\documentclass[prb,twocolumn,superscriptaddress,citeautoscript,showpacs,amsart]{revtex4}

\usepackage{graphicx}
\usepackage{multirow}
\usepackage{color}
\usepackage{bm}
\usepackage{times}
\usepackage{amsmath,bm,amsfonts}
\usepackage{dcolumn}
\usepackage{graphicx}
\usepackage{latexsym}

\usepackage{ulem} 
\usepackage{braket}

\usepackage{cancel}

\def\ket#1{|#1\rangle }
\def\bra#1{\langle #1 |}

\def\d{\partial}

\begin{document}
\title{
Stiefel-Whitney classes and topological phases in band theory
}

\author{Junyeong \surname{Ahn}}
\affiliation{Department of Physics and Astronomy, Seoul National University, Seoul 08826, Korea}

\affiliation{Center for Correlated Electron Systems, Institute for Basic Science (IBS), Seoul 08826, Korea}

\affiliation{Center for Theoretical Physics (CTP), Seoul National University, Seoul 08826, Korea}

\author{Sungjoon \surname{Park}}
\affiliation{Department of Physics and Astronomy, Seoul National University, Seoul 08826, Korea}

\affiliation{Center for Correlated Electron Systems, Institute for Basic Science (IBS), Seoul 08826, Korea}

\affiliation{Center for Theoretical Physics (CTP), Seoul National University, Seoul 08826, Korea}

\author{Dongwook \surname{Kim}}
\affiliation{Department of Physics, Sungkyunkwan University, Suwon 16419, Korea}

\author{Youngkuk \surname{Kim}}
\affiliation{Department of Physics, Sungkyunkwan University, Suwon 16419, Korea}

\author{Bohm-Jung \surname{Yang}}
\email{bjyang@snu.ac.kr}
\affiliation{Department of Physics and Astronomy, Seoul National University, Seoul 08826, Korea}

\affiliation{Center for Correlated Electron Systems, Institute for Basic Science (IBS), Seoul 08826, Korea}

\affiliation{Center for Theoretical Physics (CTP), Seoul National University, Seoul 08826, Korea}

\date{\today}

\begin{abstract}
In this article, we review the recent progress in the study of topological phases in systems with space-time inversion symmetry $I_{\text{ST}}$. $I_{\text{ST}}$ is an anti-unitary symmetry which is local in momentum space and satisfies $I_{\text{ST}}^2=1$ such as $PT$ or $C_{2}T$ symmetry where $P$, $T$, $C_2$ indicate inversion, time-reversal, and two-fold rotation symmetries, respectively. Under $I_{\text{ST}}$, the Hamiltonian and the Bloch wave function can be constrained to be real-valued, which makes the Berry curvature and the Chern number to vanish. In this class of systems, gapped band structures of real wave functions can be topologically distinguished by Stiefel-Whitney numbers instead. The first and second Stiefel-Whitney numbers $w_1$ and $w_2$, respectively, are the corresponding invariants in 1D and 2D, which are equivalent to the quantized Berry phase and the $Z_2$ monopole charge, respectively. We first describe the topological phases characterized by the first Stiefel-Whitney number, including 1D topological insulators with quantized charge polarization, 2D Dirac semimetals, and 3D nodal line semimetals. Next we review how the second Stiefel-Whitney class characterizes the 3D  nodal line semimetals carrying a $Z_{2}$ monopole charge. In particular, we explain how the second Stiefel-Whitney number $w_2$, the $Z_{2}$ monopole charge, and the linking number between nodal lines are related. Finally, we review the properties of 2D and 3D topological insulators characterized by the nontrivial second Stiefel Whitney class. 
\end{abstract}

\maketitle








\section{Introduction}

The energy band structure of a periodic crystal consists of a mapping from a crystal momentum ${\bf k}$ to the Bloch Hamiltonian $H({\bf k})$, which is generally complex-valued. Gapped band structures of insulators can be topologically distinguished by the equivalence class of $H({\bf k})$~\cite{hasan2010colloquium}. That is, two gapped band structures are topological distinct if one cannot be smoothly deformed to the other while keeping the energy gap. In two dimensions (2D), the gapped band structures of a complex Hamiltonian are distinguished by an integer topological invariant, called the Chern number $c_1$. The Chern invariant characterizes the equivalence classes of fiber bundles associated with the complex Bloch wave functions $\ket{u({\bf k})}$, which is nothing but the first Chern class. The Chern number $c_1$ can be expressed in terms of the Berry connection ${\bf A}_{m}({\bf k})=\bra{u_{m}({\bf k})}i\nabla\ket{u_{m}({\bf k})}$ and the Berry curvature $F_{m}({\bf k})=\partial_{k_x}A_{m,y}({\bf k})-\partial_{k_y}A_{m,x}({\bf k})$ where $m=1,2,...,N_{\text{occ}}$ with $N_{\text{occ}}$ denoting the number of occupied bands. $c_1$ is given by the integral of $F_{m}({\bf k})$ on a 2D closed manifold $\cal{M}$ summed over all the occupied bands as
\begin{align}
c_1=\sum_{m=1}^{N_{\text{occ}}}\int_{\cal{M}}\frac{d^2k}{2\pi}F_{m}({\bf k}).
\end{align}
When the full 2D Brillouin zone torus is considered for the integration, the corresponding insulator with a nonzero $c_1$ is a quantum Hall insulator exhibiting a quantized anomalous Hall conductivity $\sigma_{xy}=\frac{e^2}{h}c_1$. The integration can also be performed over a 2D closed subspace of a 3D Brillouin zone enclosing Weyl points. In this case, a nonzero $c_1$ corresponds to the total chiral charge of the enclosed Weyl points~\cite{armitage2018weyl,murakami2007phase,wan2011topological}.

On the other hand, when the system satisfies a certain symmetry condition, the corresponding Bloch wave functions can be real-valued. Here the relevant symmetry is so-called the space-time inversion symmetry $I_{\text{ST}}:(t,{\bf r})\rightarrow(-t,-{\bf r})$, which inverts  the sign of both time $t$ and spatial coordinates ${\bf r}$~\cite{fang2015new,ahn2017unconventional,ahn2018band}. $I_{\text{ST}}$ is an antiunitary symmetry operator that is local in momentum space and satisfies $I_{\text{ST}}^{2}=1$. For instance, the combination of spatial inversion $P$ and time-reversal $T$ can be used to define $I_{\text{ST}}=PT$ in systems with negligible spin-orbit coupling in any dimension. In the case of 2D systems, the combined symmetry $C_{2z}T$, where $C_{2z}$ denotes two-fold rotation about the $z$-axis, can also play the role of $I_{\text{ST}}$, irrespective of the presence or the absence of spin-orbit coupling~\cite{fang2015new}. Since $I_{\text{ST}}$ can always be represented by $I_{\text{ST}}=K$ with the complex conjugation operator $K$ under a suitable basis choice~\cite{fang2015new}, the invariance of the Hamiltonian $H({\bf k})$ under $I_{\text{ST}}$ imposes the reality condition to $H({\bf k})$ and $\ket{u({\bf k})}$ as
\begin{align}\label{eq:realitycondition}
&I_{\text{ST}}H({\bf k})I_{\text{ST}}^{-1}=H^{*}({\bf k})=H({\bf k}),
\nonumber\\
&I_{\text{ST}}\ket{u({\bf k})}=\ket{u({\bf k})}^{*}=\ket{u({\bf k})}.
\end{align}
Since real wave functions have zero Berry curvature at every momentum ${\bf k}$, all gapped real band structures have zero Chern number so that they are all topologically trivial in view of the first Chern class. 

\begin{table}[t]
\begin{tabular}{c | c | c }
\hline
\hline
$d$ & Physical invariant & Mathematical invariant\\
\hline
\hline
$d=1$ & Quantized Berry phase & First Stiefel-Whitney number ($w_1$)
\\
$d=2$ & $Z_2$ monopole charge & Second Stiefel-Whitney number ($w_2$)
\\
\hline \hline
\end{tabular}
\caption{
The correspondence between the physics terminology and Stiefel-Whitney (Stiefel-Whitney) class in systems with space-time inversion symmetry.
$d$ indicates the spatial dimension.
}
\end{table}\label{table:1}

In fact, the gapped band structures of real Hamiltonians are topologically distinguished by different topological invariants, so-called the Stiefel-Whitney numbers~\cite{nakahara2003geometry,hatcher2003vector, hatcher2002algebraic}. The first and the second Stiefel-Whitney numbers, $w_1$ and $w_2$, respectively, are the corresponding 1D and 2D topological invariants~\cite{shiozaki2017topological,ahn2018band}.
The Stiefel Whitney numbers $w_1$, $w_2$ are rooted in the mathematical structure of real vector bundles associated with real Bloch wave functions, which is nothing but the Stiefel Whitney class, characterizing the twist of real Bloch states in momentum space. Although the concept of Stiefel-Whitney numbers is not popular in condensed matter physics, their physical implication is quite transparent. Namely, $w_1$ is equivalent to the quantized Berry phase, while $w_2$ is equivalent to the $Z_{2}$ monopole charge of a nodal line, both are well-defined in systems with space-time inversion symmetry. The correspondence between Stiefel-Whitney number and the relevant physical invariant is summarized in Table~\ref{table:1}.


The band crossing condition changes significantly in the presence of the reality condition given in Eq.~(\ref{eq:realitycondition}).
For instance, in $PT$-symmetric systems with negligible spin-orbit coupling, since each band is non-degenerate at a generic momentum ${\bf k}$, an accidental band crossing can be described by an effective two-band Hamiltonian given by
\begin{align}\label{eqn:H_bandcrossing}
H({\bf k},m)=f_0({\bf k},m)+f_1({\bf k},m)\sigma_x+f_3({\bf k},m)\sigma_z,
\end{align}
where $\sigma_{x,y,z}$ are the Pauli matrices for the two crossing bands and $f_{0,1,3}({\bf k},m)$ are real functions of momentum $\bf{k}$ and a parameter $m$ tuning the band gap. It is worth noting that the $\sigma_y$ term vanishes due to the reality condition.
Then because closing the band gap requires only two conditions $f_{1,3}({\bf k},m)=0$ to be satisfied whereas the number of independent variables (${\bf k}$, $m$) is $d+1$ where $d$ is the spatial dimension, an accidental band crossing is possible unless $d+1<2$. This means that, in 1D, a gapped band structure is generally stable but an accidental band crossing can happen at the critical point between two gapped insulators. On the other hand, in 2D, a gapless phase with point nodes can be stabilized, and an accidental band crossing induces a transition between an insulator and a stable 2D Dirac semimetal phase. Similarly, in 3D, a semimetal phase with nodal lines can be stabilized, and an accidental band crossing mediates a transition between an insulator and a 3D nodal line semimetal. It is worth noting that the 1D insulators, the 2D Dirac points, and the 3D nodal lines mentioned above are all characterized by the first Stiefel Whitney number $w_1$, which follows from the equivalence between $w_1$ and the quantized Berry phase.

In fact, a line node of a 3D nodal line semimetal can also be characterized by another topological invariant, that is, the second Stiefel-Whitney number $w_2$. Namely, a line node of a 3D nodal line semimetal carries two $Z_2$ topological indices $w_1$ and $w_2$, and thus it is doubly charged~\cite{bzdusek2017robust}. As shown in Ref.~\onlinecite{ahn2018band}, $w_2$ is equivalent to the $Z_{2}$ monopole charge proposed in Ref.~\onlinecite{fang2015new}. In fact, a two-band description based on Eq.~(\ref{eqn:H_bandcrossing}) cannot capture $w_2$ because there should be at least two bands below the Fermi level to characterize the band topology of a nodal line carrying nonzero $Z_2$ monopole charge (monopole nodal line) at the Fermi level. It is shown that the multi-band description is required due to the intrinsic linking structure of monopole nodal line with other nodal lines below the Fermi level, which can be predicted by using the mathematical property of Stiefel Whitney classes.

The significance of $w_2$ is not limited to the characterization of monopole nodal line in 3D nodal line semimetals. When $w_2$ is computed on the 2D Brillouin zone torus, it becomes a well-defined 2D $Z_2$ topological invariant characterizing 2D insulators in the absence of the Berry phase~\cite{ahn2018band}. Thus one can classify $I_{\text{ST}}$-symmetric 2D insulators into topologically trivial insulators with $w_2=0$ and topologically nontrival insulators with $w_2=1$, dubbed the 2D Stiefel Whitney insulator (Stiefel-Whitney insulator)~\cite{ahn2018band,ahn2018failure}. Contrary to the 2D quantum Hall insulator that has stable band topology and 1D chiral edge states, the 2D Stiefel-Whitney insulator is an obstructed atomic insulator, which can support zero-dimensional corner charges in the presence of additional chiral symmetry~\cite{ahn2018band,ahn2018failure,po2018fragile,
cano2018topology,bouhon2018wilson,wang2018higher,bradlyn2018disconnected,
song2018all,po2018faithful,liu2018shift}. The 2D Stiefel-Whitney insulator can also be used as a basic building block for novel 3D topological insulators such as 3D weak and strong Stiefel-Whitney insulators as shown in Ref.~\onlinecite{ahn2018band,ahn2018higher}. Table II summarizes the correspondence between the first Chern class and the second Stiefel Whitney class.

\begin{table}[t]
\begin{tabular}{ c | c }
\hline
\hline
First Chern class ($c_1)$ & Second Stiefel-Whitney class ($w_2$)\\
\hline
\hline
2D complex wave function & 2D real wave function
\\
Chiral charge of Weyl points & $Z_2$ monopole charge of nodal lines
\\
Quantum Hall insulator & Stiefel-Whitney insulator
\\
\hline \hline
\end{tabular}
\caption{
Correspondence between the first Chern class and the second Stiefel Whitney class.
}
\end{table}\label{table:2}

The rest of the paper is organized as follows. The mathematical definition of the first and second Stiefel Whitney classes are given in Sec.II. Sec.III describes the topological phases characterizied by the first Stiefel-Whitney number. Basically, $w_1$ characterizes 1D insulators with quantized charge polarization, and 2D or 3D semimetals with point or line nodes. Sec.IV describes the relation between the second Stiefel Whitney class and the $Z_2$ monopole charges of nodal line semimetals. Using the mathematical properties of the Stiefel Whitney classes, we show the intrinsic linking structure of semimetals with $Z_{2}$ monopole charge. Also, we explain how $w_2$ can be calculated from Wilson loop spectra. Topological insulators characterized by $w_2$ are described in Sec.V. There we introduce the definition of 2D Stiefel-Whitney insulator, 3D weak Stiefel-Whitney insulator, and 3D strong Stiefel-Whitney insulator, and elaborate their topological properties. We conclude our review in Sec.VI with the discussion of candidate materials and possible issues for future studies. 

\section{Stiefel-Whitney classes}

Here we provide the mathematical definition of Stiefel-Whitney classes, which basically indicate the topological obstruction to defining real wave functions smoothly over a closed manifold.

\subsection{The first Stiefel-Whitney class}

The first Stiefel-Whitney class measures the orientability of real occupied wave functions over a closed 1D manifold. Namely, the real occupied wave functions are orientable (non-orientable) when $w_1=0$ ($w_1=1$). 

In general, the orientation of a real vector space refers to the choice of an ordered basis.
Any two ordered bases are related to each other by a unique nonsingular linear transformation.
When the determinant of the transformation matrix is positive (negative), we say that the bases have the same (different) orientation.
After choosing an ordered reference basis $\{v_1,v_2,...\}$, the orientation of another basis $\{u_1,u_2,...\}$ is specified to be positive (negative) when the basis has the same (different) orientation with respect to the reference basis.

Real wave functions defined on the Brillouin zone can be considered as real unit basis vectors defined at each momentum, that is, they form a structure of a real vector bundle over the Brillouin zone.
The basis can be smoothly defined locally on the manifold, but may not be smooth over a closed submanifold $\cal M$ of our interest.
We say that the real wave functions are orientable over $\cal M$ when local bases can be glued with transition functions having only positive determinant, i.e., all transition functions are orientation-preserving.
The orientable wave functions are classified into two classes with the positive and negative orientation as in the case of the real vector spaces.

Interestingly, the orientablity of real wave functions can be determined by the Berry phase computed in a smooth complex gauge, such that $w_1=1$ ($w_1=0$) indicates that the relevant wave functions carry $\pi$ ($0$) Berry phase. Namely, the first Stiefel-Whitney number defined in a real basis is equivalent to the well-known quantized Berry phase defined in a smooth complex basis.
This correspondence can be seen by investigating how the $\pi$ Berry phase computed with the smooth complex state $\ket{u_{n\bf k}}$ affects the real state $\ket{\tilde{u}_{n\bf k}}$ connected by a gauge transformation as
\begin{align}
\ket{\tilde{u}_{n\bf k}}=g_{mn}({\bf k})\ket{u_{m\bf k}},
\end{align}
where $g$ is the gauge transformation matrix.
In order to make the state $\ket{u_{n\bf k}}$ real, the $\pi$ Berry phase should be eliminated by a local phase rotation because the diagonal components of the Berry connection are zero when the state is real. Then we have
\begin{align}
0=\int^{2\pi}_0dk\;{\rm Tr}\tilde{A}=\int^{2\pi}_0dk({\rm Tr}A+i\nabla_{k}\log \det g),
\end{align}
where $\tilde{\bf A}_{mn}=\braket{\tilde{u}_{m\bf k}|i\nabla_{\bf k}|\tilde{u}_{n\bf k}}$ and ${\bf A}_{mn}=\braket{u_{m\bf k}|i\nabla_{\bf k}|u_{n\bf k}}$.
Integrating the $\log \det g$ term gives
\begin{align}\label{eq:w1andBerryphase}
\frac{\det g(2\pi)}{\det g(0)}=\exp \left[i \int^{2\pi}_0dk\;{\rm Tr}A\right].
\end{align}
Moreover, at the Brillouin zone boundary, since the smooth complex state fulfills $\ket{u_{n(2\pi)}}=\ket{u_{n(0)}}$, the real state satisfies
\begin{align}
\ket{\tilde{u}_{n(2\pi)}}=[g^{-1}(0)g({2\pi})]_{mn}\ket{\tilde{u}_{m(0)}},
\end{align}
which, together with Eq.~(\ref{eq:w1andBerryphase}), shows that the total Berry phase basically determines the determinant of the transition function for $\ket{\tilde{u}_{nk}}$ at the Brillouin zone boundary.
Namely, when the total Berry phase is $\pi$, the real state $\ket{\tilde{u}_{nk}}$ requires an orientation-reversal between $k=2\pi$ and $k=0$. Therefore we conclude that the first Stiefel-Whitney number $w_1$ for a closed curve $C$ in the Brillouin zone is given by
\begin{align}
w_1|_C=\frac{1}{\pi}\oint_C d{\bf k}\cdot {\rm Tr}{\bf A}({\bf k})~~~ (\text{mod}~~2),
\end{align}
where ${\bf A}({\bf k})$ is the Berry connection calculated in a smooth complex gauge.

In fact, the first Stiefel-Whitney number $w_1$ determines the orientability of real states even in higher dimensions~\cite{hatcher2003vector}.
From the analysis in 1D, we find
\begin{align}
\frac{\det g({\bf q})}{\det g({\bf p})}=\exp \left[i \int^{\bf q}_{\bf p}d{\bf k}\cdot {\rm Tr}{\bf A}({\bf k})\right].
\end{align}
Let us note that $\det g$ is globally smooth when the Berry phase is zero over every closed cycle.
Otherwise, $\det g$ becomes discontinuous at some points so that the real states are non-orientable as in the 1D case.
Thus, real states are orientable over an arbitrary dimensional closed manifold $\cal M$ if and only if the total Berry phase, which is calculated in a smooth complex gauge, is trivial for any 1D closed loop in $\cal M$.

\subsection{The second Stiefel-Whitney class}

The second Stiefel-Whitney class describes whether a spin (or pin) structure is allowed or not for given real wave functions defined on a 2D closed manifold $\cal{M}$. If $w_2=0$ ($w_2=1$), a spin or pin structure is allowed (forbidden). Below we give a more formal definition of the second Stiefel-Whitney number $w_2$.

Let us consider real occupied states $\ket{u_{m\bf k}}$ on $\cal{M}$. Then we introduce a covering of $\cal{M}$ whose geometric structure is topologically equivalent to $\cal{M}$. For given two open covers (or patches) $A$ and $B$, one can find smooth real wave functions $\ket{u^A_{m\bf k}}$ and $\ket{u^B_{m\bf k}}$ defined in each open cover, respectively. In the overlapping region $A\cap B$, a transition function $t^{AB}_{mn}({\bf k})$ is defined as
\begin{align}\label{eqn:tAB}
\ket{u^B_{n\bf k}}=t^{AB}_{mn}({\bf k})\ket{u^A_{m\bf k}}.
\end{align}
For convenience, here we assume that the occupied wave functions are orientable over $\cal{M}$ so that $t^{AB}_{mn}({\bf k})\in \text{SO}(N_{\text{occ}})$. Then, since the double covering of $\text{SO}(N_{\text{occ}})$ is $\text{Spin}(N_{\text{occ}})$, the problem reduces to whether the lifting of the transition function from $t\in\text{SO}(N_{\text{occ}})$ to $\tilde{t}\in\text{Spin}(N_{\text{occ}})$ is allowed or not. If such a lifting is allowed consistently over $\cal{M}$, one can say that the spin structure exists, and thus $w_2=0$. In contrast, if such a lifting is not allowed, the spin structure does not exist, and $w_2=1$. The extension to the case of non-orientable manifolds is also straightforward as shown in Ref.~\onlinecite{ahn2018band}.

In general, the transition functions should satisfy the following consistency conditions~\cite{nakahara2003geometry},
\begin{align}
\label{consistency}
t^{AB}_{\bf k}t^{BA}_{\bf k}=1,
\end{align}
for ${\bf k}\in A\cap B$ and
\begin{align}
\label{triple_consistency}
t^{AB}_{\bf k}t^{BC}_{\bf k}t^{CA}_{\bf k}=1,
\end{align}
for ${\bf k}\in A\cap B\cap C$, where $A$, $B$, and $C$ are arbitrary patches.
The transition functions defined in Eq.~(\ref{eqn:tAB}) satisfy these consistency conditions automatically.

However, when we consider the lifting of transition functions to the double covering group at all overlapping regions in $\cal{M}$, the consistency conditions are not automatically satisfied everywhere.
Let us write $I$ and $-I$ to denote the $0$ and $2\pi$ rotation in the double covering group.
In general, after the lift $t^{AB}\rightarrow \tilde{t}^{AB}$, the lifted transition functions satisfy 
\begin{align}
\tilde{t}^{AB}_{\bf k}\tilde{t}^{BA}_{\bf k}=\pm I
\end{align}
for ${\bf k}\in A\cap B$ and
\begin{align}
f^{ABC}_{\bf k}
\equiv \tilde{t}^{AB}_{\bf k}\tilde{t}^{BC}_{\bf k}\tilde{t}^{CA}_{\bf k}
=\pm I,
\end{align}
for ${\bf k}\in A\cap B\cap C$.
The sign can be either $+$ or $-$ because both $I$ and $-I$ are projected to 1 via a two-to-one mapping from $\text{Spin}(N_{\text{occ}})$ to $\text{SO}(N_{\text{occ}})$.
$f^{ABC}_{\bf k}$ is gauge-invariant as one can see from the transformation of the lifted transition functions 
$\tilde{t}^{AB}_{\bf k}\rightarrow (\tilde{g}^{A}_{\bf k})^{-1}\tilde{t}^{AB}_{\bf k}\tilde{g}^B_{\bf k}$ 
under $\ket{u^A_{n\bf k}}\rightarrow g^A_{mn \bf k}\ket{u^A_{m\bf k}}$, where $\tilde{g}$ is a lift of $g$.
Also, $f^{ABC}_{\bf k}$ has a unique value at each triple overlap, because it is fully symmetric with respect to the permutation of $A$, $B$, $C$, and is independent of ${\bf k}$ within a triple overlap.

Let us now examine the case where a lift that satisfies the consistency conditions can be found. In general, there is no obstruction for the first consistency condition in Eq.~(\ref{consistency}) whereas the second consistency condition in Eq.~(\ref{triple_consistency}) depends on $w_2$. In fact, $w_2$ is defined as
\begin{align}
\label{Cech_Stiefel-Whitney2}
(-I)^{w_2}=\prod_{A\cap B\cap C}f^{ABC},
\end{align}
where the product is over all triple overlaps in $\cal{M}$. One can see that transition functions cannot be lifted to their double covering group when $w_2=1$ modulo two, because, in this case, there is at least one triple overlap $A\cap B\cap C$ where $f^{ABC}=-I$ violating the consistency condition.
This indicates that the obstruction to the existence of a spin structure is dictated by the $Z_2$ invariant $w_2$.

\begin{figure}[t!]
\includegraphics[width=8.5cm]{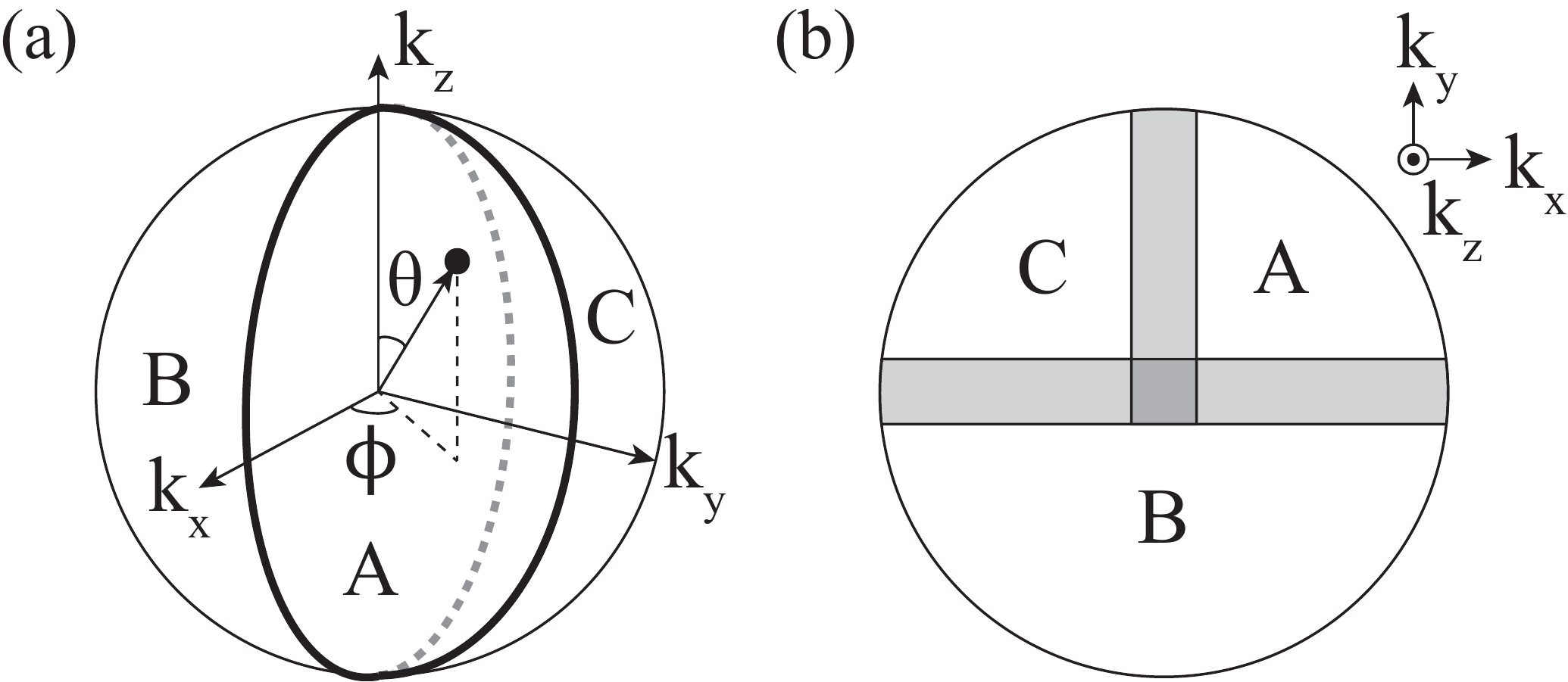}
\caption{Three patches covering a sphere.
(a) Orthographic view.
$\phi$ and $\theta$ are the azimuthal and polar angles.
$B$ and $A$ ($C$) overlap at $\phi=0$ ($\phi=\pi$), and $A$ and $C$ overlap at $\phi=\pi/2$.
The north pole at $\theta=0$ and the south pole at $\theta=\pi$ are triple overlaps.
(b) Top view.
Overlapping regions are exaggerated for clarity.
Figures are adoped from the supplemental materials in Ref.~\onlinecite{ahn2018band}.
}
\label{sphere_patch}
\end{figure}

For instance, let us illustrate how $w_2$ is defined on a spherical manifold, which is directly relevant to the $Z_{2}$ monopole charge of a nodal line.
First we consider three patches $A$, $B$, and $C$ covering a sphere shown in Fig.~\ref{sphere_patch}.
In the spherical coordinates $(\phi,\theta)$, there are three overlaps $A\cap B$, $A\cap C$, and $B\cap C$ at $\phi=0$, $\phi=\pi/2$, and $\phi=\pi$, respectively.
We restrict all transition functions on the overlaps to $\text{SO}(N_{\text{occ}})$, which is possible because every loop on a sphere is contractible to a point such that the first Stiefel-Whitney number is trivial.
Then
\begin{align}\label{eqn:W2_sphere}
(-I)^{w_2}
&=f^{ABC}(0)f^{ABC}(\pi),
\end{align}
where $0$ and $\pi$ denotes the polar angle $\theta$.

Now let us define
\begin{align}
\tilde{W}(\theta)=\tilde{t}^{AB}(\theta)\tilde{t}^{BC}(\theta)\tilde{t}^{CA}(\theta),
\end{align}
where we omit $\phi$ in the argument of transition functions because they are uniquely specified by the overlapping region.
$\tilde{W}(\theta)$ is smooth for $0<\theta<\pi$ because $\tilde{t}$ is smooth within an overlap.
$\tilde{W}(0)=f^{ABC}(0)=\pm I$, and $\tilde{W}(\pi)=f^{ABC}(\pi)=\pm I$.
Then we see that $w_2=1$ modulo two when the image of the map $\tilde{W}:[0,\pi]\rightarrow \text{Spin}(N_{\text{occ}})$ is an arc connecting $I$ and $-I$, whereas $w_2=0$ when the image is a closed loop containing $I$ or $-I$.
Next, we project $\tilde{W}$ to $W$ by using the two-to-one map $\text{Spin}(N_{\text{occ}})\rightarrow \text{SO}(N_{\text{occ}})$.
We have
\begin{align}
\label{winding_sphere}
W(\theta)=t^{AB}(\theta)t^{BC}(\theta)t^{CA}(\theta),
\end{align}
which is smooth for $0<\theta<\pi$, and $W(0)=W(\pi)=1$. Under this projection, an arc connecting $I$ and $-I$ projects to a loop winding the non-contractible cycle an odd number of times, whereas a closed loop projects to a contractible loop or a non-contractible loop winding the non-contractible cycles an even number of times~\cite{kirby1991pin}. As a result, the second Stiefel-Whitney number is given by the winding number of $W({\theta})$ modulo two. This relation between $w_2$ and the parity of the winding number of $W({\theta})$ provides the correspondence between the second Stiefel-Whitney number and the $Z_2$ monopole charge of a nodal line enclosed by the sphere, which is further discussed in Sec.IV.

\section{First Stiefel-Whitney class and topological phases}

\begin{table}[t]
\begin{tabular}{ c | c | c }
\hline
\hline
Topological phase & $w_1$ & $w_2$\\
\hline
\hline
Insulator & 1D TI with $P_1=1/2$ & 2D SWI
\\
Semimetal & 2D DSM or 3D NLSM & 3D monopole NLSM
\\
\hline \hline
\end{tabular}
\caption{
Comparison of the topological phases characterized by the first Stiefel-Whitney number $w_1$ and the second Stiefel-Whitney number $w_2$. Here TI (SWI) indicates a topological insulator (Stiefel Whitney insulator). DSM (NLSM) denotes Dirac semimetal (nodal line semimetal). A monopole nodal line indicates a nodal line with $Z_2$ monopole charge.
}
\end{table}\label{table:3}

Here we describe the topological phases in $I_{\text{ST}}$-symmetric systems characterized by the first Stiefel-Whitney number $w_1$. In 1D, $w_1$ computed over the full Brillouin zone is the bulk topological invariant of insulators with quantized charge polarization. In 2D (3D), $w_1$ is defined on a closed loop enclosing a point (line) nodes, and plays the role of a topological charge carried by the node. In all these cases, the low-energy properties of the system can be generally described by the $2\times2$ effective Hamiltonian in Eq.~(\ref{eqn:H_bandcrossing}) and the corresponding band crossing occurs when
\begin{align}\label{eqn:bandcrossingcondition}
f_{1}({\bf k},m)=f_{3}({\bf k},m)=0.
\end{align}

\subsection{1D topological insulator: SSH model in a real basis}
\begin{figure}[t!]
\includegraphics[width=8.5cm]{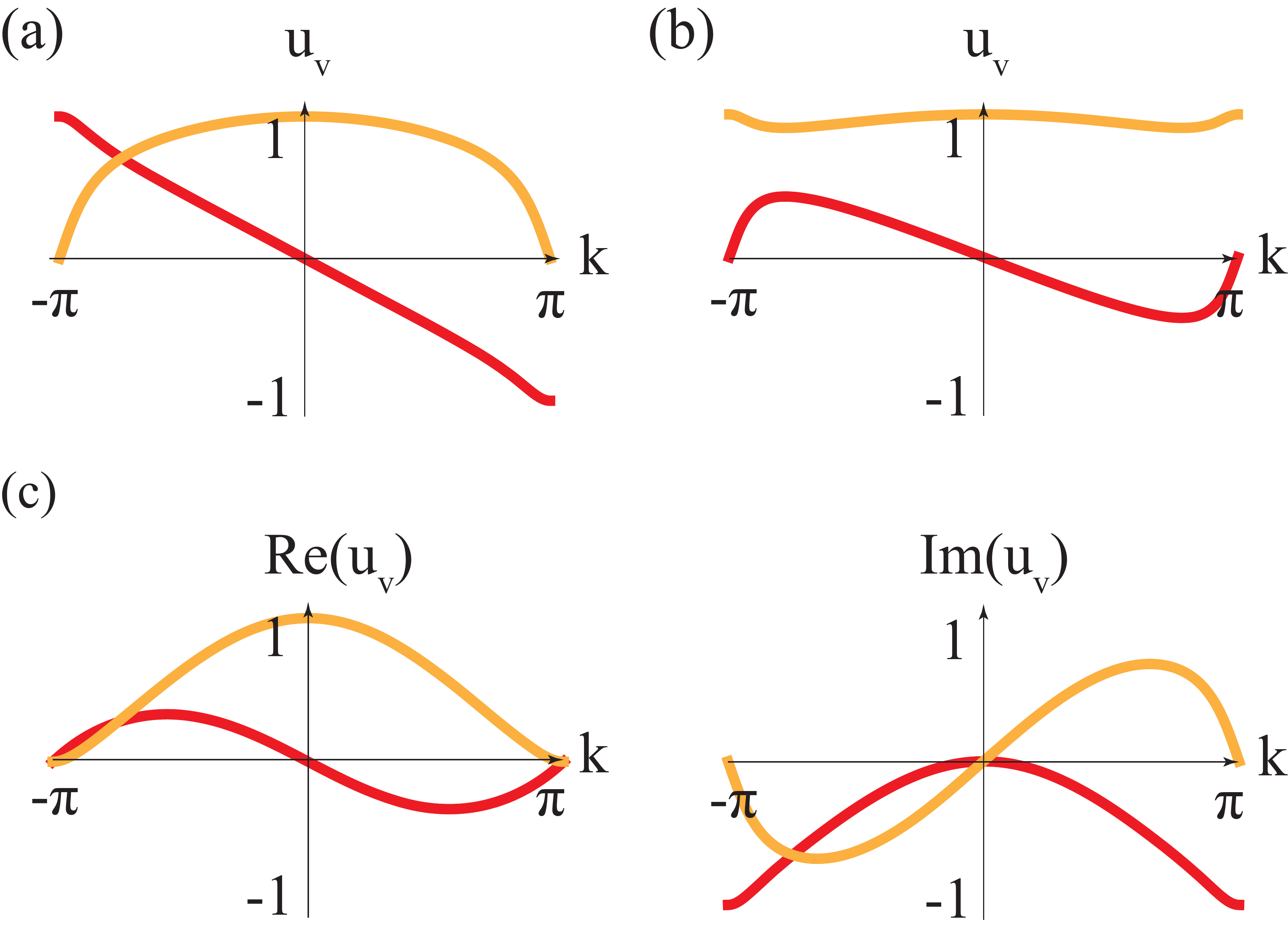}
\caption{The amplitudes of the occupied state of the SSH model.
Red and orange lines show the first and second components of the occupied state in Eq.~(\ref{SSH_VB}).
(a,b) Under reality condition. The occupied state is anti-periodic for
(a) $t=0.7$
whereas it is periodic for
(b) $t=1.3$.
(c,d) In a smooth complex gauge with $\phi_v(k)=k/2$. The occupied state is smooth both for (c) $|t|<1$ and (d) $|t|>1$.
Figures are adoped from the supplemental materials in Ref.~\onlinecite{ahn2018band}.
}
\label{orientation}
\end{figure}
In 1D, since there is only one momentum variable $k$ for a given tuning parameter $m$, the gap-closing condition in Eq.~(\ref{eqn:bandcrossingcondition}) cannot be satisfied in general, and thus an insulating phase becomes stable. However, when $k$ and $m$ are varied simultaneously, one can find a unique solution of Eq.~(\ref{eqn:bandcrossingcondition}) that corresponds to the critical point between two gapped insulators.

To describe the relation between the topological property of an insulator and $w_1$, let us consider 1D insulators described by the Su-Schuriffer-Heeger (SSH) model~\cite{su1979solitons}. Although the properties of the corresponding topological insulator are already well-known, here we describe the SSH model by using a real basis and illustrate the relevant topological property in the context of the first Stiefel-Whitney number. The SSH Hamiltonian is given by
\begin{align}
H_{\rm SSH}=\sin k \sigma_x+(t+\cos k)\sigma_z,
\end{align}
where the Pauli matrices $\sigma_{x,y,z}$ indicate the two sublattice sites within a unit cell. This Hamiltonian is symmetric under $PT=K$ so that the bulk charge polarization $P_1$ is quantized into either $P_1=0$ or $P_1=1/2$ modulo 1.
It is well-known that this system describes a topologically trivial insulator  with $P_1=0$ when $|t|> 1$ and a topologically nontrivial insulator (TI) with $P_1=1/2$ when $|t|<1$.
Let us see how the topology manifests in the wave function for the occupied state, which is given by
\begin{align}
\label{SSH_VB}
\ket{u_v}=\frac{e^{i\phi_v(k)}}{N(k)}
\begin{pmatrix}
\sin k\\
t+\cos k-\sqrt{(t+\cos k)^2+\sin^2k}
\end{pmatrix},
\end{align}
where $\phi_v(k)$ is an arbitrary overall phase factor and $N(k)$ is a positive normalization factor.

First, let us impose the reality condition on $\ket{u_v}$ over the whole 1D Brillouin zone by choosing the gauge $e^{i\phi_v(k)}=1$.
Fig.~\ref{orientation}(a) and (b) show the amplitudes of the two components of $\ket{u_v}$ along the Brillouin zone when $|t|<1$ an $|t|>1$, respectively. When $|t|>1$, the real wave function is smooth and continuous over the entire 1D Brillouin zone, and thus $w_1=0$. In contrast, the real wave function is discontinuous at the Brillouin zone boundary at $k=\pm\pi$ when $|t|<1$, although it is smooth over $-\pi<k<\pi$. Then at the boundaries $k=\pm \pi$, the wave function should be glued with an orientation-reversing transition function. This demonstrates that the occupied state is non-orientable, and thus $w_1=1$.

On the other hand, if the reality condition is relaxed by choosing $\phi_v(k)=k/2$, the occupied state becomes globally smooth in both $|t|<1$ and $|t|>1$ cases. [See Fig.~\ref{orientation}(c), (d).] However, the discontinuity of the real wave function for $|t|<1$ manifests as a $\pi$ Berry phase of the corresponding smooth complex wave function.
In this smooth complex gauge, one can easily show that $A=\braket{u_v|i\nabla_k|u_v}=1/2$  and $P_1=\int^{\pi}_{-\pi}dk A/(2\pi)=1/2$ when $|t|<1$. Similarly, we find $P_1=0$ when $|t|>1$. This example clearly demonstrates that $w_1=1$ ($w_1=0$) in a real gauge is equivalent to the $\pi$ Berry phase (0 Berry phase) in a smooth complex gauge.

\subsection{2D Dirac semimetal}
In 2D, the gap-closing condition in Eq.~(\ref{eqn:bandcrossingcondition}) can be satisfied by tuning two momenta $k_x$ and $k_y$ for given $m$, which indicates that a semimetal with point nodes is stable in general. When $m$ is also treated as a tuning parameter, the number of  variables ($k_{x}$, $k_y$, $m$) becomes larger than the number of equations in Eq.~(\ref{eqn:bandcrossingcondition}), which predicts a line of gapless solutions in the 3D parameter space ($k_{x}$, $k_y$, $m$). Physically, this indicates an insulator-semimetal transition acrosss the band crossing generating a stable 2D Dirac semimetal phase.

The stability of the Dirac points in the resulting Dirac semimetal phase can be understood in the following way. Since the Berry curvature $F_{xy}({\bf k})$ transforms to $-F_{xy}({\bf k})$ under $I_{\text{ST}}$, $F_{xy}({\bf k})$ vanishes locally unless there is a singular gapless point, which leads to the quantization of $\pi$ Berry phase around a 2D Dirac point, which, at the same time, guarantees its stability~\cite{fang2015new,C2T_Serbyn,C2T_Hsieh,C2T_Fang}. In terms of real wave functions, discontinuous real wave functions along a loop encircling an odd number of Dirac points cannot be adiabatically connected to smooth real wave functions along a loop encircling an even number of Dirac points, which again indicates the stability of Dirac points.

\begin{figure}[t]
\centering
\includegraphics[width=8.5cm]{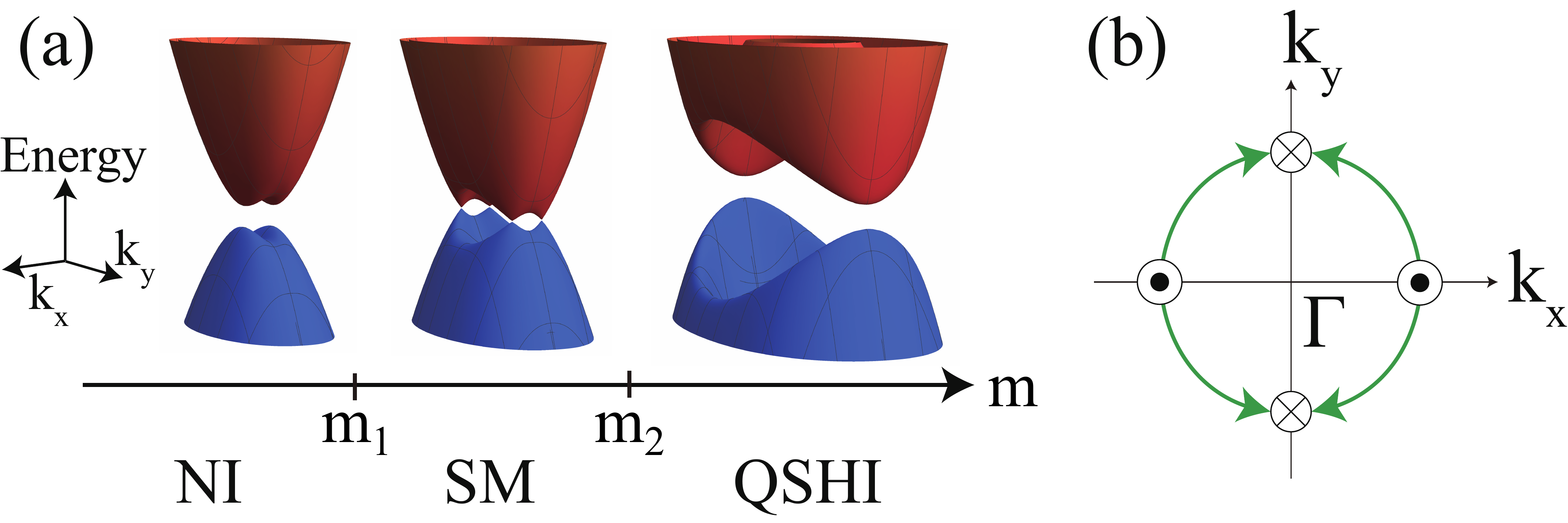}
\caption{(Color online)
(a) Evolution of the band structure of a 2D noncentrosymmetric system with $I_{\text{ST}}$ symmetry as a tuning parameter $m$ increases. 
(b) Trajectories of 2D Weyl points in the intermediate semimetal phase as $m$ increases. 
$\odot$ and $\otimes$ are the locations where pair creation and pair annihilation happen.
Figures are adoped from the supplemental materials in Ref.~\onlinecite{ahn2017unconventional}. 
}
\label{fig:2dTPT}
\end{figure}

After the discovery of graphene, it has been well-known that a stable 2D Dirac point can exist in $PT$-symmetric 2D spinless fermion systems. However, in this system, it is also known that the Dirac point becomes unstable once spin-orbit coupling is turned on. Therefore finding a Dirac point stable in the presence of spin-orbit coupling was an interesting open question. One interesting idea proposed in Ref.~\onlinecite{YoungKane2D,WiederKane2D} is to use the nonsymmophic crystalline symmetry which protects four-fold degenerate Dirac points at the corners of the Brillouin zone in 2D centrosymmetric systems. 

On the other hand, in the case of 2D noncentrosymmetric systems, $C_{2z}T$ symmetry can protect Dirac points with two-fold degeneracy whose location can be anywhere in the Brillouin zone~\cite{fang2015new}. One can call a gap-closing point with two-fold degeneracy as a 2D Weyl point, which is distinguished from four-fold degenerate Dirac points in centrosymmetric systems. As long as the inversion is broken, and thus the spin splitting occurs at a generic momentum, the above description in Eq.~(\ref{eqn:H_bandcrossing}) based on $2\times2$ matrix is still valid in $C_{2z}T$ symmetric noncentrosymmetric systems with spin-orbit coupling. In fact, near the critical point $m=m_{c1}$ for the insulator-semimetal transition, the Hamiltonian can generally be written as
\begin{align}\label{eqn:criticalHamiltonian}
H({\bf q})=(Aq_{x}^{2}+m_{c1}-m)\sigma_{x}+vq_{y}\sigma_{z},
\end{align}
which describes a gapped insulator (a 2D semimetal) when $m<m_{c1}$ ($m>m_{c1}$) assuming $A>0$. 
Due to $T$ symmetry, accidental band crossing happens at two momenta $\pm{\bf k}$. Since two Weyl points are created at each band crossing point, the semimetal phase has four Weyl points in total. 
Moreover, when $m$ becomes larger than $m_{c1}$, four Weyl points migrate in momentum space, and eventually, they are annihilated pairwise at another critical point $m=m_{c2}$.  Interestingly, when pair creation/pair annihilation is accompanied by a partner-switching between the Weyl point pairs, the two gapped phases mediated by the Weyl semimetal should have distinct topological property~\cite{ahn2017unconventional}. Namely, if one gapped phase is a normal insulator, the other one should be a 2D quantum spin Hall insulator as shown in Fig.~\ref{fig:2dTPT}. This theory predicts that the topological phase transition in HgTe/CdTe heterostructrue~\cite{QSH_theory,QSH_review} should be mediated by a 2D Weyl semimetal phase inbetween. Indeed, the pair creation of 2D Weyl points across an insulator-semimetal transition was experimentally observed in few-layer black phosphorus under vertical electric field~\cite{KSKim_PRL}.
The condition for accidental band crossing in 2D noncentrosymmetric systems was systematically classified by considering 2D layer group in Ref.~\onlinecite{ParkYang2D}.

\subsection{3D nodal line semimetals}
In 3D, the gap-closing condition in Eq.~(\ref{eqn:bandcrossingcondition}) can be satisfied by tuning three momenta $k_x$, $k_y$, $k_z$ for given $m$, which indicates that a semimetal with line nodes is stable in general. When $m$ is also treated as a tuning parameter, since the number of variables ($k_{x}$, $k_y$, $k_z$, $m$) is four while the number of equations in Eq.~(\ref{eqn:bandcrossingcondition}) is two, a 2D sheet of gapless solutions exists in the 4D parameter space of ($k_{x}$, $k_y$, $k_z$, $m$). This indicates an insulator-semimetal transition acrosss the band crossing generating a stable 3D nodal line (NL) semimetal. As in the case of 2D Dirac semimetals discussed above, the stability of nodal lines in this class of nodal line semimetals (NLSMs) is guaranteed by $w_1$ or $\pi$ Berry phase defined on a closed loop encircling a line node. NLSMs in $I_{\text{ST}}$-symmetric systems were proposed in various materials~\cite{NLSM1,NLSM2,NLSM3,NLSM4,NLSM5,NLSM6}. Recent developments in the study of NLSMs are extensively reviewed in Ref.~\onlinecite{NLSM_review} where NLSM whose band degeneracy is protected by other symmetries such as mirror or nonsymmorphic symmetries are also systematically reviewed.

\section{Second Stiefel-Whitney class and 3D nodal line semimetals with monopole charge}

Here we describe the properties of topological NLSM in 3D whose nontrivial topology is characterized by the second Stiefel-Whitney class. We first show that the $Z_2$ monopole charge of a NL defined on the wrapping sphere is equivalent to the second Stiefel-Whitney number $w_2$. Based on this equivalence, we apply the mathematical property of the second Stiefel-Whitney class to the problem of 3D nodal line semimetals, and predict the fundamental topological properties of the nodal line semimetals with $Z_2$ monopole charges.

\subsection{Second Stiefel-Whitney number and $Z_2$ monopole charge of nodal lines}
\begin{figure}[t!]
\includegraphics[width=8.5cm]{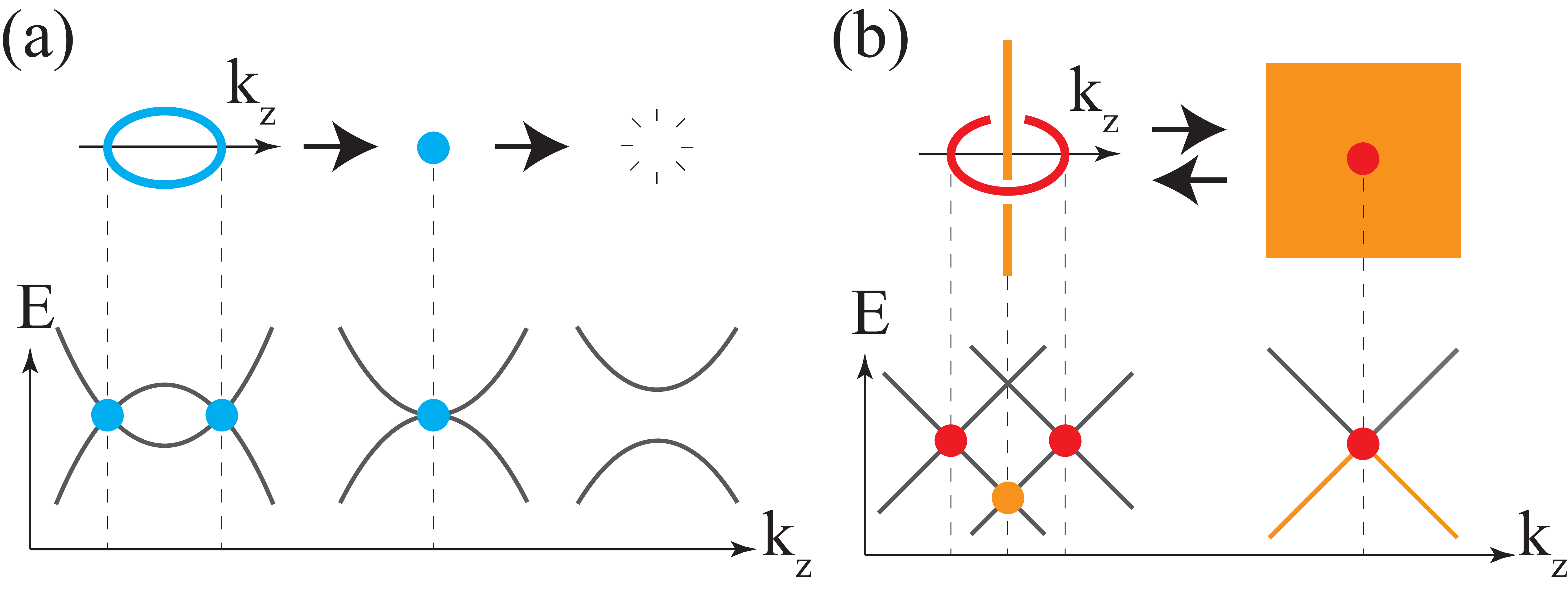}
\caption{Stabilty of nodal lines in 3D $PT$-symmetric spinless fermion systems.
(a) A $Z_2$ trivial nodal line can be gapped via band re-inversion.
(b) A nodal line carrying $Z_2$ monopole charge cannot be gapped alone.
}
\label{fig:3d_bandcrossing}
\end{figure}

In 3D $PT$-symmetric spinless fermion systems, an accidental band crossing described by the $2\times 2$ effective Hamiltonian in Eq.~(\ref{eqn:H_bandcrossing}) generates a nodal line that is locally stable due to $\pi$ Berry phase as discussed before. However, even in the presence of $\pi$ Berry phase, it is still allowed to deform a single nodal loop into a point node, which eventually disappears leading to a gapped insulator as shown in Fig.~\ref{fig:3d_bandcrossing}(a). Since an annihilation of a single nodal line is allowed, such a nodal line obviously carries a zero monopole charge. 

However, recently it has been proposed that there is another type of nodal lines carrying $Z_{2}$ monopole charge (monopole nodal line)~\cite{Fang_Z2}. A single monopole nodal line cannot be gapped and thus stable. To annihilate a monopole nodal line, one needs to introduce another monopole nodal line, and then the nodal line pair can be pair annihilated.
To describe a monopole nodal line, one needs to go beyond the two-band description given in Eq.~(\ref{eqn:H_bandcrossing}) and consider at least a four-band Hamiltonian. For instance, let us consider the following real four-band Hamiltonian
\begin{align}
\label{nlsm}
H({\bf k})=k_x\Gamma_1+k_y\Gamma_2+k_z\Gamma_3+m\Gamma_{15},
\end{align}
where $\Gamma_{(1,2,3)}=(\sigma_x,\tau_y\sigma_y,\sigma_z)$, $\Gamma_{15}=\tau_z\sigma_z$, and $\tau_{x,y,z}$ and $\sigma_{x,y,z}$ are the Pauli matrices.
The energy eigenvalues are $E=\pm\sqrt{k_x^2+\left(\rho\pm |m|\right)^2}$ where $\rho=\sqrt{k_y^2+k_z^2}$.
One can see that the conduction and the valence bands touch along the closed loop satisfying $k_x=0$ and $\rho=|m|$. This nodal line carries a nonzero monopole charge, which can be confirmed by directly computing the monopole charge. Let us note that once the reality condition is imposed, only three Gamma matrices, $\Gamma_{1,2,3}$ shown above, six matrices $\Gamma_{ab}$ $(a=1,2,3; b=4,5)$, and $I$ can appear in $H(\bf{k})$. 
The presence of three real Gamma matrices, which mutually anticommutes, indicates that a 3D massless Dirac fermion can exist in this system~\cite{Morimoto_Z2,zhao2017pt}.
The Dirac point is stable against the gap opening because the mass terms associated with $\Gamma_{4,5}$ are forbidden. However, the six real matrices $\Gamma_{ab}$ can deform the Dirac point into a nodal line while keeping the system gapless.

Moreover, there is another intriguing nodal structure in the bands that the Hamiltonian in Eq.~(\ref{nlsm}) describes. That is, the occupied bands also cross and form another nodal line at $\rho=0$ (NL$^*$), which is linked with the monopole nodal line~\cite{ahn2018band}.
Because of this linking, the monopole nodal line is stable and distinct from trivial NLs.
As $m\rightarrow 0$, the linking requires that the monopole nodal line becomes a four-fold degenerate Dirac point as shown in Fig.~\ref{fig:3d_bandcrossing} (b).
If $m$ becomes finite after its sign-reversal, the size of the monopole nodal line increases again. A single monopole nodal line cannot be gapped and only a pair of monopole nodal lines can be created or annihilated across the band inversion.

\begin{figure}[t!]
\includegraphics[width=8.5cm]{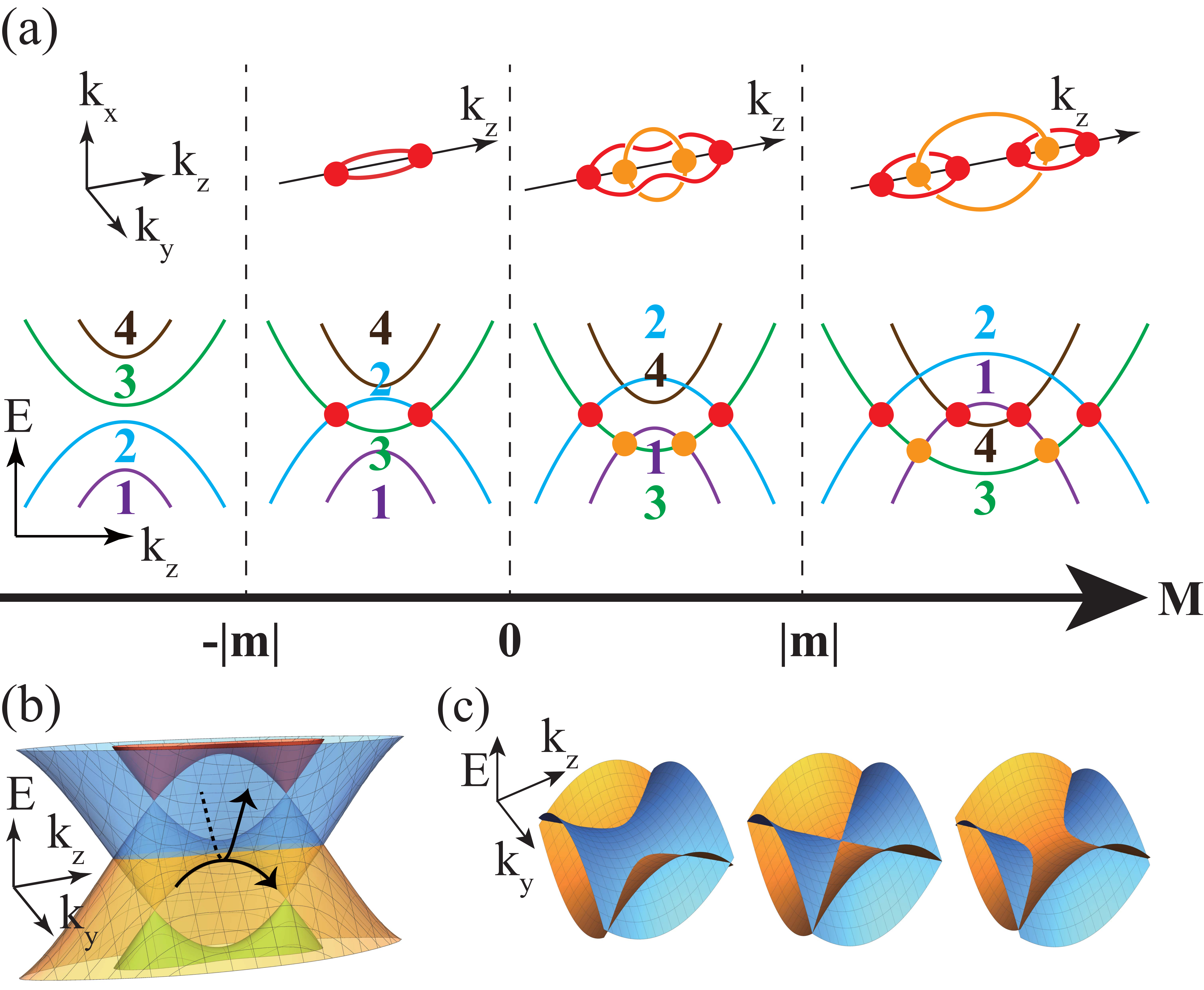}
\caption{Pair creation of nodal lines carrying $Z_{2}$ monopole charge (monopole nodal lines) via a double band inversion (DBI).
(a) Evolution of band structure during DBI.
Red (Orange) points and lines indicate the crossing between the conduction and valence bands (two occupied bands).
(b) Saddle-shaped band structure when $0<M<|m|$.
(c) Change in nodal line structure when two saddle-shaped bands cross.
Figures are adoped from the supplemental materials in Ref.~\onlinecite{ahn2018band}.
}
\label{saddle_inversion}
\end{figure}

Now let us illustrate the mechanism for the pair creation of monopole nodal lines, which is a sequence of band inversions described in Fig.~\ref{saddle_inversion}(a), dubbed {\it a double band inversion} (DBI).
For concreteness, we describe a DBI by using the Hamiltonian in Eq.~(\ref{nlsm}) after the replacement $k_z\rightarrow|{\bf k}|^2-M$.
The evolution of the band structure during the DBI is illustrated in Fig.~\ref{saddle_inversion} (a) as a function of the parameter $M$.
As $M$ is increased from $M<-|m|$, the first band inversion occurs at $M=-|m|$ between the top valence and bottom conduction bands, and it creates a trivial NL protected by $\pi$ Berry phase.
Then, the inversion at $M=0$ between the two occupied (unoccupied) bands generates another NL below (above) the Fermi level, which we call NL$^*$.
After this band inversion, the band structure near ${\bf k}=0$ develops saddle-shape around the Fermi energy as shown in Fig.~\ref{saddle_inversion}(b). 
Another consecutive band inversion at $M=|m|$ between the two saddle-shaped bands induces a Lifshitz transition as shown in Fig.~\ref{saddle_inversion}(c), during which the trivial NL splits into two monopole nodal lines, which are linked by the NL$^*$ that are formed from the occupied bands.
During DBI, each occupied (unoccupied) band crosses both of two unoccupied (occupied) bands, which is the reason why the minimal number of bands required to create a monopole nodal line is four.

Now let us discuss the equivalence between the second Stiefel-Whitney number $w_2$ and the $Z_2$ monopole charge of a nodal line. The equivalence follows from the fact that the nontrivial $Z_2$ monopole charge forbids the existence of the spin structure on a sphere enclosing the nodal line.
Let us first briefly review the idea of the $Z_2$ monopole charge that is defined over a sphere enclosing a nodal line~\cite{Fang_Z2}.
For this, we divide the wrapping sphere into two patches, each covering the northern ($N$) or the southern ($S$) hemispheres which overlap along the equator. $\ket{u^N_n({\bf k})}$ and $\ket{u^S_n({\bf k})}$ are real wave functions defined smoothly on the northern and the southern hemispheres, respectively. On the overlapping region, $\ket{u^N_n({\bf k})}$ and $\ket{u^S_n({\bf k})}$ are related by a transition function $t^{NS}$ in a way that $\ket{u^S_n({\bf k})}=t^{NS}_{mn}({\bf k})\ket{u^N_m({\bf k})}$ for ${\bf k}\in N\cap S$.
The $Z_2$ monopole charge is defined by the winding number of the transition function~\cite{Fang_Z2}.
We restrict the transition function to $\text{SO}(N_{\text{occ}})$, which is possible because every loop on a sphere is contractible to a point, that is, the corresponding first Stiefel-Whitney number is trivial.
Then we see that the winding number of $t^{NS}$ along a loop in $N\cap S$ gives a $Z_2$ number because $\pi_1(\text{SO}(N_{\text{occ}}))=Z_2$ for $N_{\text{occ}}>2$.
This $Z_2$ number is nothing but the $Z_2$ monopole charge.
When the number of occupied bands is two, the winding number is integer-valued because $\pi_1(\text{SO}(2))=Z$.
In this case, the $Z_2$ monopole charge is given by the parity of the winding number.

Interestingly, this $Z_2$ monopole charge also characterizes the obstruction to having a spin structure over the wrapping sphere.
For simplicity, let us take a gauge where the transition function $t^{NS}({\bf k})$ is an identity at some ${\bf k}={\bf k}_0\in N\cap S$.
Then, the transition function undergoes a $4\pi \mathcal{N}$ rotation ($2\pi(2\mathcal{N}+1)$ rotation) for an integer $\mathcal{N}$ along a loop containing ${\bf k}_0$ in $N\cap S$ when the $Z_2$ monopole charge is trivial (nontrivial). This is because the homotopy group $\pi_1(\text{SO}(N_{\text{occ}}))$ is generated by the paths representing $2\pi\mathcal{N}$ rotation~\cite{kirby1991pin,prasolov2006elements} and the $Z_{2}$ monopole charge indicates the parity of the relevant winding number.
While the $2\pi$ rotation and the identity are identical in $\text{SO}(N_{\text{occ}})$, they are not identical in ${\rm }\text{Spin}(N_{\text{occ}})$.
Therefore, the transition function is well-defined over the overlap $N\cap S$ only as a $\text{SO}(N_{\text{occ}})$ element when the $Z_2$ monopole charge is nontrivial.
On the other hand, no obstruction arises when the $Z_2$ monopole charge is trivial because a $4\pi$ rotation is identical to the identity element even as a $\text{Spin}(N_{\text{occ}})$ element. Thus, the $Z_2$ monopole charge is identical to the second Stiefel-Whitney number over the enclosing sphere. This equivalence can be proved more rigorously by using the mathematical definition given in Eq.~(\ref{Cech_Stiefel-Whitney2}), (\ref{eqn:W2_sphere}) after suitable deformation of the patches and related transition functions as shown in Ref.~\onlinecite{ahn2018band}

\subsection{Whitney sum formula and linking of nodal lines}

Since the equivalence between the $Z_{2}$ monopole charge and the second Stiefel-Whitney number $w_2$ has been established, one can use the mathematical properties of Stiefel-Whitney numbers to understand the physical properties of monopole nodal lines. One important mathematical property of Stiefel-Whitney numbers is the so-called Whitney sum formula, which provides the rule for determining the total second Stiefel-Whitney number of blocks of bands from $w_1$ and $w_2$ of each block.

To explain the Whitney sum formula, let us suppose that the set of the occupied bands ${\cal B}$ can be decomposed into a direct sum of $n$ subsets.
\begin{align}
{\cal B}=\bigoplus_{i}{\cal B}_i=B_1\oplus {\cal B}_2...\oplus {\cal B}_n.
\end{align}
Here the direct sum indicates that one can find a basis in which the transition function can be block-diagonal over all the overlapping regions $A\cap B$ between two patches $A$ and $B$. For instance, the Whitney sum formula can be applied to the gapped band structure between different blocks of bands. However, even if different blocks are not gapped, the formula is valid as long as the transition function has a block-diagonal form under a suitable basis.
Explicitly, the Whitney sum formula is given by~\cite{hatcher2003vector,chooquet-bruhat2000analysis}
\begin{align}
\label{Whitney_sum}
w_2\left(\oplus_{i}{\cal B}_i\right)=\sum_{i}w_2({\cal B}_i)+\sum_{i<j}\frac{1}{\pi^2}\oint_{\cal M} d{\bf S}\cdot {\bf A}_{i}\times {\bf A}_{j},
\end{align}
where ${\cal M}$ indicates a closed 2D manifold and ${\bf A}_i=\sum_{n\in {\cal B}_i}\braket{u_{n\bf k}|i\nabla_{\bf k}|u_{n\bf k}}$ is the Berry connection for the $i$-th block which is calculated in a smooth complex gauge.
On a torus, the second term in Eq.~(\ref{Whitney_sum}) can be expressed as
\begin{align}
\label{Kunneth}
\oint_{T^2} d{\bf S}\cdot {\bf A}_{i}\times {\bf A}_{j}
&=\Phi_{i,\phi}\Phi_{j,\theta}-\Phi_{i,\theta}\Phi_{j,\phi}
\end{align}
where $\Phi_{i,\psi}=\oint A_{i,\psi}d\psi$ is the Berry phase of the $i$-th block computed on the non-contractible cycle along the $\psi=\theta$ or $\phi$ direction.
Because the Berry phase $\Phi$ in a smooth complex gauge is the first Stiefel-Whitney number $w_1$ in a real gauge, the Berry phase $\Phi_{i,\psi}$ in Eq.~(\ref{Kunneth}) can be replaced by $\pi w_1({\cal B}_i)$ leading to
\begin{align}
\label{Whitney_torus}
w_2\left(\oplus_{i}{\cal B}_i\right)&=\sum_{i}w_2({\cal B}_i)
\nonumber\\
&+\sum_{i<j}\left[w_1^{\phi}({\cal B}_i)w_1^{\theta}({\cal B}_j)-w_1^{\theta}({\cal B}_i)w_1^{\phi}({\cal B}_j)\right],
\end{align}
which is valid on a torus ${\cal M}=T^{2}$.

One important physical implication of the Whitney sum formula given in Eq.~(\ref{Whitney_torus}) is that it reveals the linking structure of a monopole nodal line at the Fermi level with other nodal lines below the Fermi level. Since $w_2$ is equivalent to the $Z_2$ monopole charge of a nodal line, the relation between $w_2$ and the linking number of a nodal line also provides the equivalence of the $Z_2$ monopole charge of a nodal line to its linking number modulo two.

To show the relation between the $w_2$ and the linking number, let us consider a sphere enclosing a nodal line $\gamma_1$ at the Fermi level. $w_2$ defined on the sphere is identical to the $Z_2$ monopole charge of the nodal line $\gamma_1$. Now we smoothly deform the sphere into a torus while keeping the energy gap finite during the deformation. Then, $w_2$ computed on the sphere and the torus should be the same. If the torus $T^2$ wrapping $\gamma_1$ is thin enough, all the occupied bands can become non-degenerate everywhere on the torus. Then, because the $w_2$ of a non-degenerate band is zero, the Whitney sum formula becomes
\begin{align}
\label{Stiefel-Whitney_torus}
w_{2}=&\sum_{n<m}\frac{1}{\pi^2}\oint_{T^2} d{\bf S}\cdot {\bf A}_{n}\times {\bf A}_{m},
\end{align}
where ${\bf A}_n=\braket{u_{n\bf k}|i\nabla_{\bf k}|u_{n\bf k}}$ is the Berry connection for the $n$th topmost occupied band, and $n$ and $m$  run over the occupied bands.

\begin{figure}[t!]
\includegraphics[width=5cm]{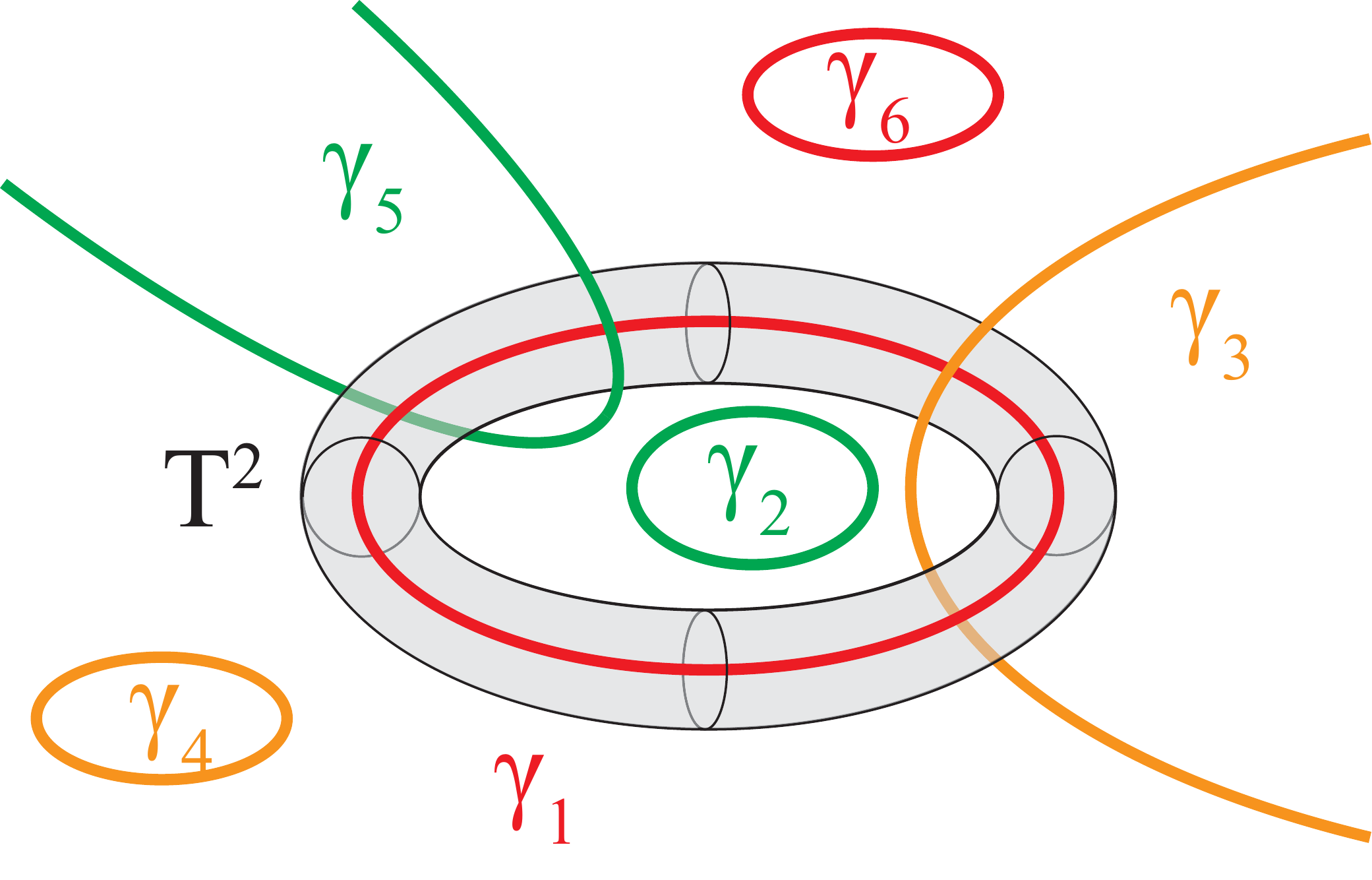}
\caption{A torus $T^2$ wrapping the nodal line $\gamma_1$ in the Brillouin zone.
$T^2$ is thin enough so that it does not intersect any other band degeneracies.
Red lines ($\gamma_1$ and $\gamma_6$) are lines of touching between the conduction and valence band.
Orange ($\gamma_3$ and $\gamma_4$) and green ($\gamma_2$ and $\gamma_5$) lines are lines of touching between the first and the second topmost occupied band and between the second and the third topmost occupied band, respectively.
Only the linking between $\gamma_1$ and $\gamma_3$ is protected as it generates the nontrivial second Stiefel-Whitney number on $T^2$, whereas the linking between $\gamma_1$ and $\gamma_5$ is not protected [See Eq.~(\ref{Stiefel-Whitney=linking})].
Figures are adoped from the supplemental materials in Ref.~\onlinecite{ahn2018band}.
}
\label{wrapping_torus}
\end{figure}

By noting that the quantization condition of the Berry phase, $\oint_C A_n=\pi$ or 0, resembles the Ampere's law in electromagnetics, one can solve the equation and get a solution analogous to the Bio-Savart law.
Explicitly, let us start from the differential form representing the quantization of the Berry phase, which is similar to the differential form of the Ampere's law,
\begin{align}
\nabla_{\bf k}\times{\bf A}_n({\bf k})
=\sum_iI_n^i\oint_{\gamma_i} d{\bf k}_{i} \delta^{3}({\bf k}-{\bf k}_{i}),
\end{align}
which gives
\begin{align}
{\bf A}_n({\bf k})
&=\frac{1}{4\pi}\sum_{i}\oint_{\gamma_i}\frac{I^i_nd{\bf k}_i\times ({\bf k}-{\bf k}_i)}{|{\bf k}-{\bf k}_i|^3},
\end{align}
where $\gamma_i$'s are lines of band touching points, and $I^i_n=\pi$ if the line $\gamma_i$ generates the Berry phase on $n$th band whereas $I^i_n=0$ otherwise.

By using this formula of the Berry connection, we have
\begin{align}
\label{linking_derivation}
&\frac{1}{\pi^2}\oint_{T^2} d{\bf S}\cdot {\bf A}_{n}\times {\bf A}_{m}
=\sum_{\gamma_j}\left(\delta_{1n}\frac{I_m^j}{\pi}-\delta_{1m}\frac{I_n^j}{\pi}\right){\rm Lk}(\gamma_1,\gamma_j),
\end{align}
where
\begin{align}
{\rm Lk}(\gamma_i,\gamma_j)
=\frac{1}{4\pi}\oint_{\gamma_i}\oint_{\gamma_j}\frac{d{\bf k}_i\times d{\bf k}_j\cdot ({\bf k}_i-{\bf k}_j)}{|{\bf k}_i-{\bf k}_j|^3},
\end{align}
is the Gauss' linking integral of the closed lines $\gamma_i$ and $\gamma_j$.
This eventually leads to
\begin{align}
\label{Stiefel-Whitney=linking}
w_2(T^2)
&=\sum_{\tilde{\gamma}_j}{\rm Lk}(\gamma_1,\tilde{\gamma}_j),
\end{align}
where $\tilde{\gamma}_j$'s are nodal lines formed between the first and second topmost occupied bands. The implication of Eq.~(\ref{Stiefel-Whitney=linking}) is clear. For convenience, let us suppose that there is one nodal line $\gamma_1$ at the Fermi level and another nodal line $\tilde{\gamma}$ below the Fermi level. If the monopole charge of $\gamma_1$ is one, $\gamma_1$ and $\tilde{\gamma}$ should be linked. On the other hand, if the monopole charge of $\gamma_1$ is zero, $\gamma_1$ and $\tilde{\gamma}$ are unlinked. This relation between the $Z_2$ monopole charge and linking number also explain why the minimal model Hamiltonian describing monopole nodal line should be at least a $4\times 4$ matrix. Since a monopole nodal line should be linked with other nodal lines below the Fermi level, there should be at least two occupied bands below the Fermi level.

\subsection{Computation of $w_2$ by using Wilson loop method}

\begin{figure}[t!]
\includegraphics[width=8.5cm]{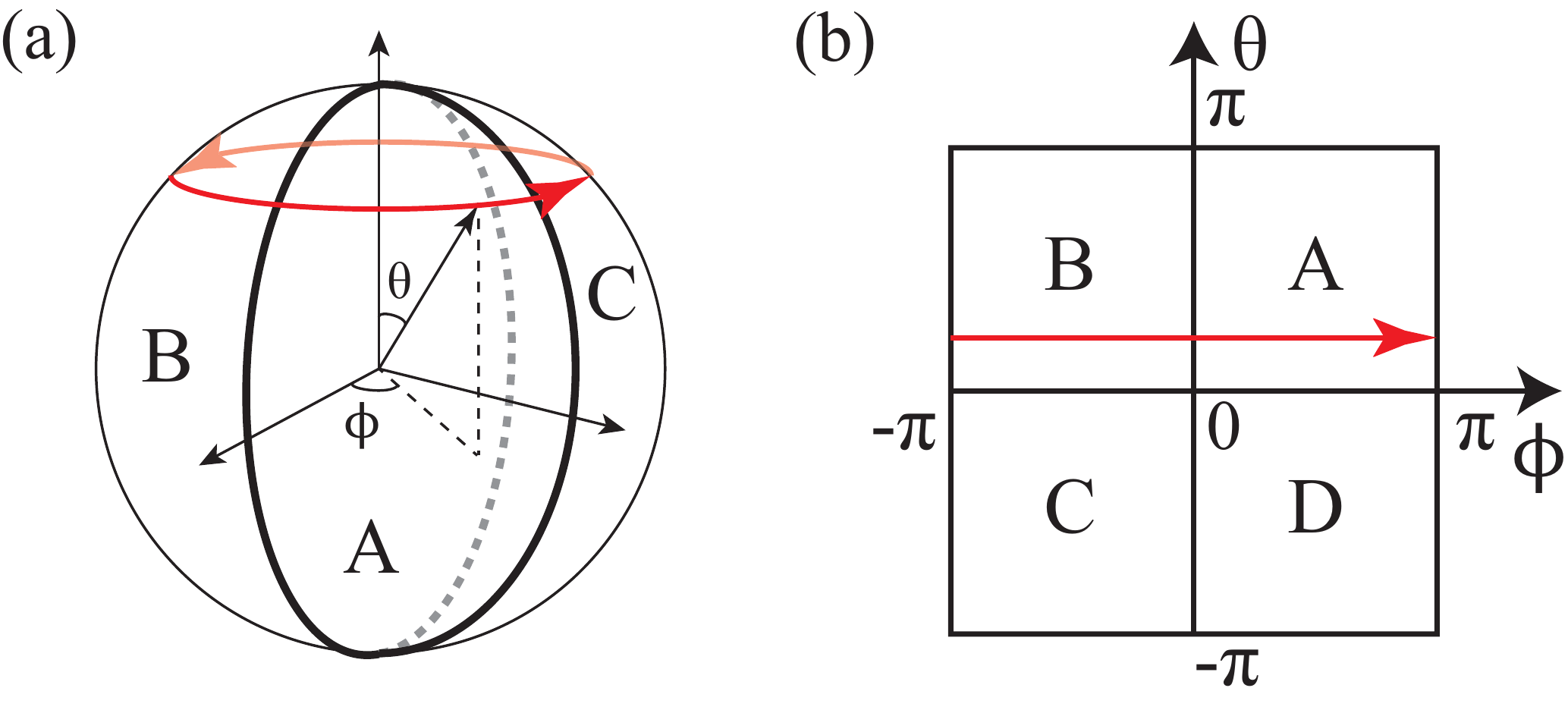}
\caption{Patches covering a sphere and a torus.
Wilson loop operators are calculated along the red arrows.
(a) Sphere covered with three patches which overlap on $\phi=0$, $\phi=\pi/2$, and $\phi=\pi$.
(b) Torus covered with four patches.
$\phi$ and $\theta$ are $2\pi$-periodic.
Transition functions are taken to be nontrivial only over $\theta=0$ and $\phi=0$ lines.
Figures are adoped from the supplemental materials in Ref.~\onlinecite{ahn2018band}.
}
\label{Wilson_patch}
\end{figure}

Let us explain how the second Stiefel-Whitney number $w_2$ can be computed by using the Wilson loop method. 
The Wilson loop operator is defined by~\cite{yu2011equivalent,alexandradinata2014wilson,alexandradinata2016topological}
\begin{align}
\label{Wilson_discrete}
W_{(\phi_0+2\pi,\theta)\leftarrow (\phi_0,\theta)}
=\lim_{N\rightarrow \infty}F_{N-1}F_{N-2}...F_{1}F_{0},
\end{align}
where $(\phi,\theta)$ parametrizes a 2D subspace of the 3D Brillouin zone or the 2D Brillouin zone itself, and $F_j$ is the overlap matrix at $\phi_j=\phi_0+2\pi j/N$ whose matrix elements are given by $[F_{j}]_{mn}=\braket{u_{m\phi_{j+1}}|u_{n\phi_j}}$, and $\phi_{N}=\phi_0$.
As shown below, the homotopy class of the Wilson loop operator gives the second Stiefel-Whitney number in a special gauge known as the parallel-transport gauge~\cite{soluyanov2012smooth}.

First, let us consider a sphere in the Brillouin zone, which is covered by three patches $A$, $B$, and $C$ whose overlaps $A\cap C$, $C\cap B$, and $B\cap A$ are at $\phi=\pi/2$, $\phi=\pi$, and $\phi=2\pi$, respectively, for all $0\le \theta\le \pi$ [See Fig.~\ref{Wilson_patch}(a)].
Here $\phi$ and $\theta$ are the azimuthal and the polar angles of the sphere.
Real occupied states $\ket{u_{n\bf k}}$ are smooth within each patch.
The Wilson loop operator $W_{0}(\theta)\equiv W_{(2\pi,\theta)\leftarrow (0,\theta)}$ is then
\begin{align}
W_{0}(\theta)
=
&\braket{u^A(0,\theta)|u^B(2\pi,\theta)}W_{(2\pi,\theta)\leftarrow (\pi,\theta)}\notag\\
&\braket{u^B(\pi,\theta)|u^C(\pi,\theta)}W_{(\pi,\theta)\leftarrow (\pi/2,\theta)}\notag\\
&\braket{u^C(\pi/2,\theta)|u^A(\pi/2,\theta)}W_{(\pi/2,\theta)\leftarrow (0,\theta)}\notag\\
=
&t^{AB}(\theta){\cal P}e^{-i\int^{2\pi}_{\pi} d\phi'  A^B_{\phi}(\theta,\phi')}\notag\\
&t^{BC}(\theta){\cal P}e^{-i\int^{\pi}_{\pi/2} d\phi'  A^C_{\phi}(\theta,\phi')}\notag\\
&t^{CA}(\theta){\cal P}e^{-i\int^{\pi/2}_{0} d\phi'  A^A_{\phi}(\theta,\phi')},
\end{align}
where we used that $W_{(\theta,\phi_2)\leftarrow (\theta,\phi_1)}={\cal P}e^{-i\int^{\phi_2}_{\phi_1} d\phi'  A_{\phi}(\theta,\phi')}$ when the states $\ket{u_{n(\phi,\theta)}}$ are smooth for $\phi_1<\phi<\phi_2$, and $A_{nm,\phi}=\braket{u_{m(\theta,\phi)}|i\d_{\phi}|u_{n(\theta,\phi)}}$ is the $\phi$ component of the Berry connection. Here ${\cal P}$ indicates that the integral is path-ordered.
If we take the parallel-transport gauge, which is defined by
\begin{align}
\label{parallel-transport_gauge_sphere}
\ket{u^A_{p,n{(\phi,\theta)}}}
&=\left[{\cal P}e^{-i\int^{\phi}_{0} d\phi'  A^A_{\phi}(\theta,\phi')}\right]_{mn}\ket{u^A_{m(\phi,\theta)}},\notag\\
\ket{u^B_{p,n{(\phi,\theta)}}}
&=\left[{\cal P}e^{-i\int^{\phi}_{\pi} d\phi'  A^B_{\phi}(\theta,\phi')}\right]_{mn}\ket{u^B_{m(\phi,\theta)}},\notag\\
\ket{u^C_{p,n{(\phi,\theta)}}}
&=\left[{\cal P}e^{-i\int^{\phi}_{\pi/2} d\phi'  A^C_{\phi}(\theta,\phi')}\right]_{mn}\ket{u^C_{m(\phi,\theta)}},
\end{align}
the Wilson loop operator is then
\begin{align}
W_{0}(\theta)
=W_{p,0}(\theta)
=t^{AB}_p(\theta)t^{BC}_p(\theta)t^{CA}_p(\theta),
\end{align}
where $W_p$ and $t_p$ are the Wilson loop operator and the transition function in the parallel-transport gauge.
It is worth noting that the Wilson loop operator in parallel-transport gauge is given by the product of all the transition functions along the $\phi$ cycle for given $\theta$, which has the same form of Eq.~(\ref{winding_sphere}). Since $W_{0}(\theta=0,\pi)=1$ due to the consistency condition at triple overlaps, the image of the map $W_{0}(\theta)$ for $\theta\in[0,\pi]$ forms a closed loop. Then the second Stiefel-Whitney number $w_2$ is given by the parity of the winding number of $W_0(\theta)$ as explained in Sec.II.B, which can be obtained gauge invariantly from its eigenvalues $\Theta(\theta)$.

\begin{figure}[t!]
\includegraphics[width=8.5cm]{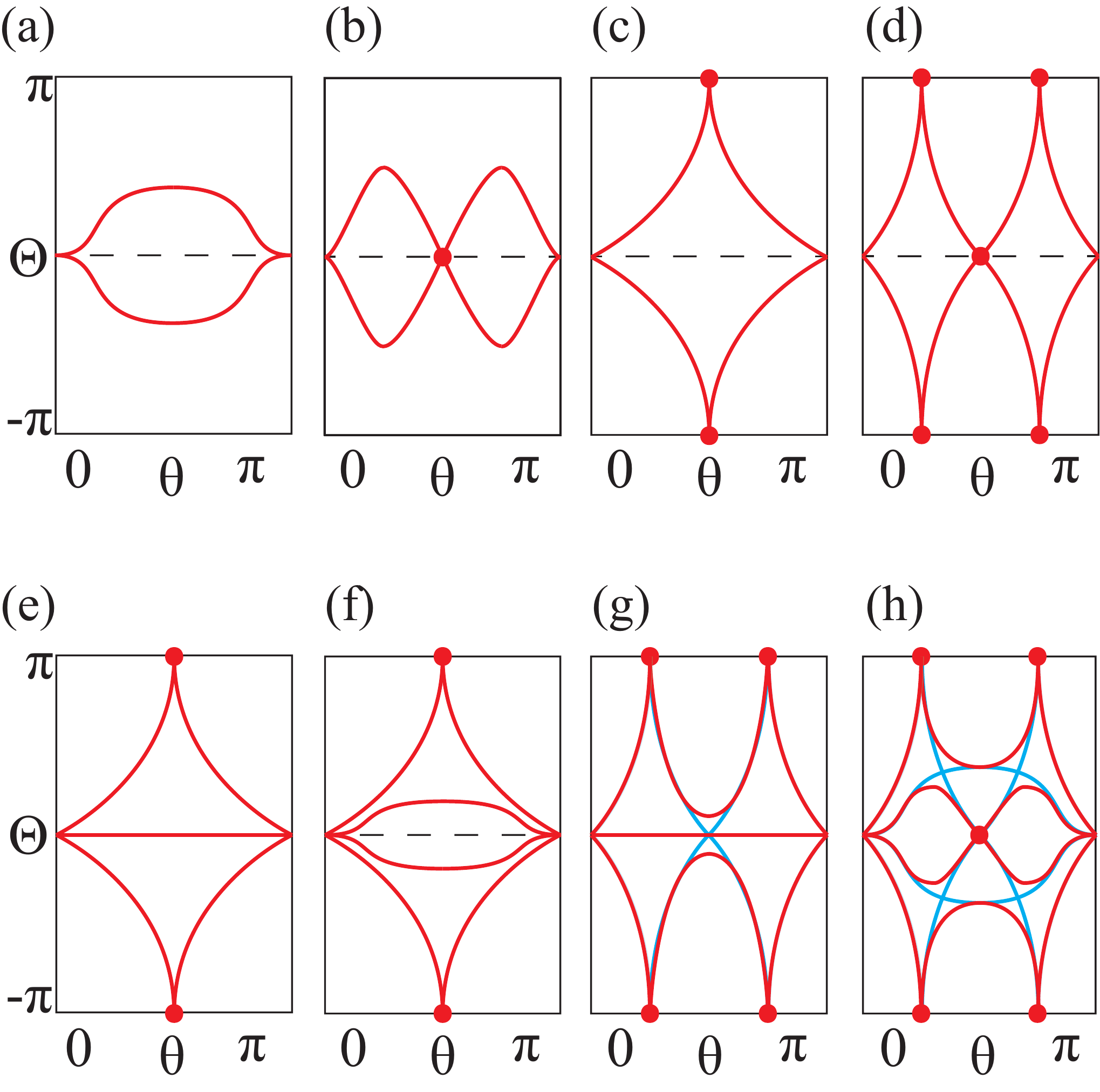}
\caption{Wilson loop spectra on a sphere.
The Wilson loop operator is calculated along the azimuthal direction with a fixed polar angle $\theta$.
(a-d) When the number of occupied bands $N_{\text{occ}}$ is two.
The winding number is (a,b) zero, (c) one, and (d) two.
(b) can be adiabatically deformed to (a) as we push the crossing point on $\Theta=0$ out of the boundary at $\theta=0$ or $\theta=\pi$.
(e-h) When $N_{\text{occ}}=3$ or $N_{\text{occ}}=4$.
(e) has a flat spectrum on $\Theta=0$ in addition to (c).
Adding a small perturbation to (e) does not deform the spectrum when $N_{\text{occ}}=3$, while it deforms the spectrum to (f) when $N_{\text{occ}}=4$.
(g) and (h) shows the $Z_2$ nature of the Wilson loop spectrum for $N_{\text{occ}}=3$ and $N_{\text{occ}}=4$, respectively.
Blue lines in (g) and (h) are obtained after adding one and two flat bands to (d), respectively, whereas red lines are the spectrum after adding a $PT$-preserving deformation which eliminates non-protected crossing points.
The crossing points on $\Theta=\pi$ can always be pair-annihilated after this deformation.
Figures are adoped from the supplemental materials in Ref.~\onlinecite{ahn2018band}.
}
\label{Wilson_sphere}
\end{figure}

Fig.~\ref{Wilson_sphere}(a-d) shows four examples of Wilson loop spectra when $N_{\text{occ}}=2$.
As Fig.~\ref{Wilson_sphere}(a) and (b) both have zero winding number, Fig.~\ref{Wilson_sphere}(a) can be smoothly deformed to Fig.~\ref{Wilson_sphere}(b), and vice versa.
It is possible as the crossing point on $\Theta=0$ can be annihilated at the boundary $\theta=0$ or $\theta=\pi$.
However, Fig.~\ref{Wilson_sphere}(b) cannot be adiabatically deformed to Fig.~\ref{Wilson_sphere}(c) or (d) because they have different winding numbers.

Let us note that the parity of the winding number, which is equivalent to the second Stiefel-Whitney number, is given by the parity of the number of crossing points on $\Theta=\pi$.
Thus, we can get $w_2$ by counting the number of the crossing points on the line.
This is also true when the number of occupied bands is larger than two because the crossing points on $\Theta=\pi$ are stable even when additional trivial bands are added.

While a single linear crossing point is locally stable on the line $\Theta=n\pi$ for any integer $n$, two linear crossing points on the same line may be pair-annihilated.
In fact, one important difference between $N_{\text{occ}}=2$ and $N_{\text{occ}}>2$ cases is that a pair annihilation, which is forbidden in the former case, is always possible in the latter case.
In the case of two occupied bands, two linear crossing points on the line $\Theta=\pi$ at $\theta=\theta_1$ and $\theta=\theta_2$ cannot be pair-annihilated if there is a linear crossing point on the $\Theta=0$ line at $\theta=\theta_0$ satisfying $\theta_1<\theta_0<\theta_2$ [See Fig.~\ref{Wilson_sphere}(d)].
This is because both of the eigenvalues are on $\Theta=0$, and no eigenvalue exists on $\Theta=\pi$ at $\theta=\theta_0$.
Accordingly, the crossing points at $\theta_1$ and $\theta_2$ on $\Theta=\pi$ cannot be pair-annihilated because they can never reach the polar angle $\theta_0$ which is between $\theta_1$ and $\theta_2$.
On the other hand, such protection mechanism does not exist when $N_{\text{occ}}>2$ and a pair annihilation is always possible on $\Theta=\pi$ as shown in Fig.~\ref{Wilson_sphere}(g,h). Hence, only the parity of the winding number is topologically meaningful when $N_{\text{occ}}>2$. This parity is the second Stiefel-Whitney number $w_2$ which can be determined by counting the number of the crossing points on $\Theta=\pi$.

For later use, let us briefly explain how the Wilson loop operator can be computed on a torus and it is related with the second Stiefel-Whitney number. Similar to the case on a sphere, the Wilson loop operator on a torus can be related to the transition function as follows.
Let us consider a torus covered with four patches shown in Fig.~\ref{Wilson_patch}(b).
We take a gauge where transition functions are trivial over $\theta=\pi$ line.
Also, we impose on $\theta=0$ that $t^{AD}$ and $t^{BC}$ are the identity (constant orientation-reversing matrices) when the occupied states are orientable (non-orientable) along $\theta$.
Then, we move to the parallel-transport gauge which is defined by
\begin{align}
\label{parallel-transport_gauge_torus}
\ket{u^{A/D}_{p,n{(\phi,\theta)}}}
&=\left[{\cal P}e^{-i\int^{\phi}_{0} d\phi'  A^{A/D}_{\phi}(\theta,\phi')}\right]_{mn}\ket{u^{A/D}_{m(\phi,\theta)}},\notag\\
\ket{u^{B/C}_{p,n{(\phi,\theta)}}}
&=\left[{\cal P}e^{-i\int^{\phi}_{\pi} d\phi'  A^{B/C}_{\phi}(\theta,\phi')}\right]_{mn}\ket{u^{B/C}_{m(\phi,\theta)}},
\end{align}
where $0\le \phi \le \pi$ in the first line, $\pi\le \phi \le 2\pi$ in the second line.
In this gauge, the Wilson loop operator is given by the product of all the transition functions existing in the $\phi$ direction for given $\theta$ as
\begin{align}
W_{0}(\theta)
=
\begin{cases}
t_p^{AB}(0,\theta)t_p^{BA}(\pi,\theta) \quad&{\rm for}\; 0\le \theta\le \pi, \\
t_p^{DC}(0,\theta)t_p^{CD}(\pi,\theta) \quad&{\rm for}\; \pi\le \theta\le 2\pi.
\end{cases}
\end{align}
Here $W_0(\theta)$ is smoothly defined in the range $0\le \theta<2\pi$, but its periodic boundary condition is nontrivial:
\begin{align}
\label{Wilson_p.c.}
W_{0}(2\pi)
&=(t_p^{AD}(0,0))^{-1}W_{0}(0)t_p^{AD}(0,0),
\end{align}
where we used that $t_p^{AD}(\phi,0)=t^{AD}(0,0)$ and $t_p^{BC}(\phi,0)=t^{BC}(\pi,0)$ are independent of $\phi$.
Let us note that since the transition function is assumed to be trivial over the $\theta=\pi$ line, $t_p^{AD}(0,0)$ determines the orientability along the $\theta$ direction. Then Eq.~(\ref{Wilson_p.c.}) means that the orientability along the $\theta$ direction constrains the Wilson loop eigenvalues along the $\theta$ direction. This information is important to understand the Wilson loop spectrum defined on the 2D Brillouin zone torus as discussed in Sec.V.A. 
The homotopy class of the Wilson loop operator $W_0(\theta)$ determines the second Stiefel-Whitney number $w_2$, gauge-invariantly from the spectrum of $W_0(\theta)$, similar to the case on a sphere. More rigorous proof for the relationship between the homotopy class of the Wilson loop operator and the mathematical definition of the second Stiefel-Whitney number on a torus is given in Ref.~\onlinecite{ahn2018band}.

\subsection{Candidate Materials}

\begin{figure}[t!]
\includegraphics[width=8.5cm]{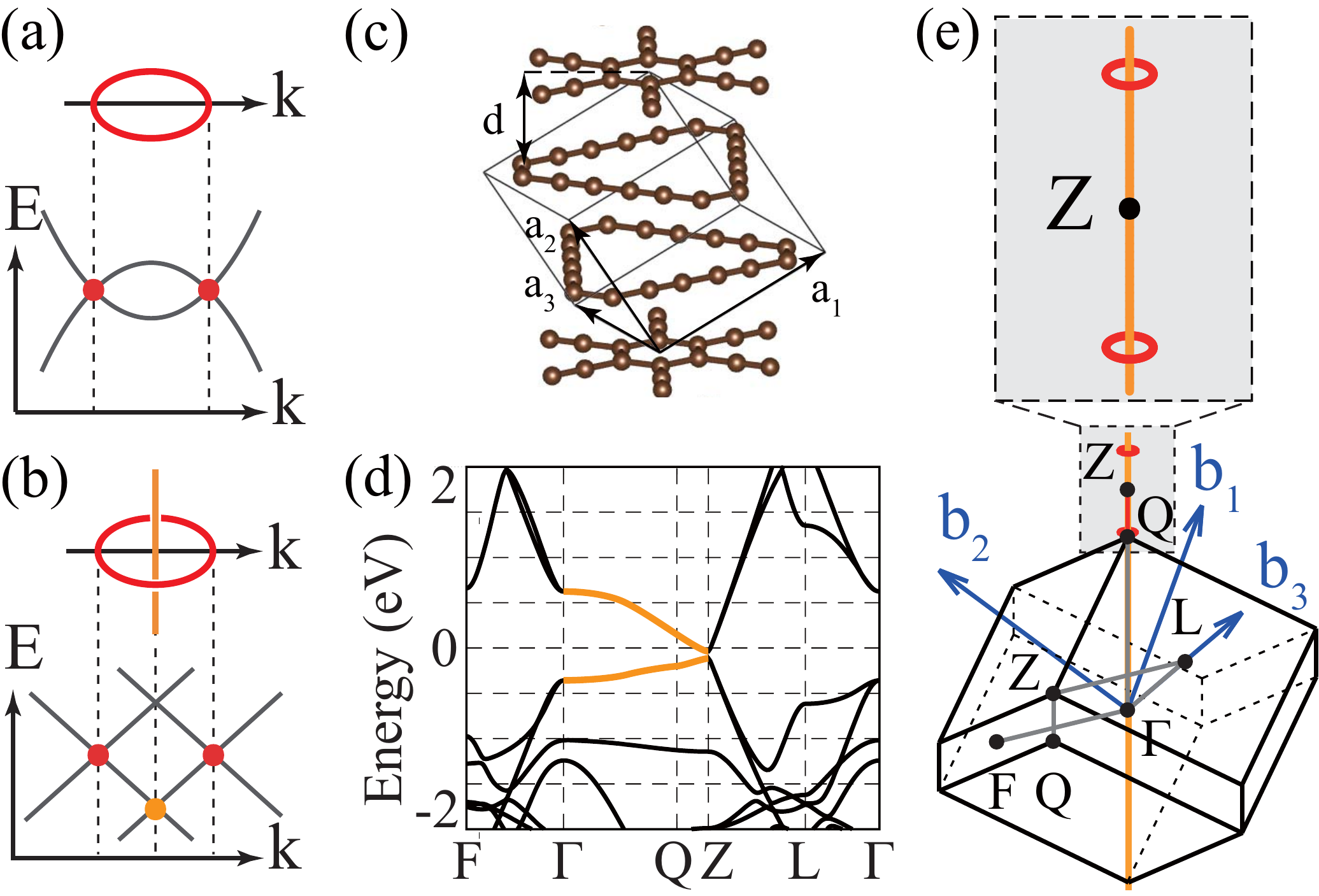}
\caption{
(a) Band structure near a nodal line with zero $Z_{2}$ monopole charge.
(b) Band structure near a nodal line carrying a unit $Z_{2}$ monopole charge (monopole nodal line) linked with another nodal line below the Fermi level ($E_{F}$).
(c) Atomic structure of ABC-stacked graphdiyne.
(d) Band structure of ABC-stacked graphdiyne where thick orange lines indicate degenerate nodal lines above and below $E_{F}$.
(e) The shape of two monopole nodal lines (red loops) at $E_{F}$ ($E=0$) linked with a nodal line below $E_{F}$ (yellow line) in ABC-stacked graphdiyne.
Figures are adoped from Ref.~\onlinecite{ahn2018band}.
}
\label{material}
\end{figure}

In Ref.~\onlinecite{ahn2018band}, based on first-principles calculations, ABC-stacked graphdiyne is proposed to realize monopole nodal lines with the linking structure. ABC-stacked graphdiyne refers to an ABC stacking of 2D graphdiyne layers composed of sp$^2$-sp carbon network of benzene rings connected by ethynyl chains. [See Fig.~\ref{material}(c).] Nomura \textit{et al.}~\cite{nomura2018three} theoretically proposed ABC-stacked graphdiyne as a nodal line semimetal, and later Ahn et al.~\cite{ahn2018band} found that it belongs to the monopole nodal line phase, characterized by the nontrivial second Stiefel-Whitney number. Consistent with Ref.~\onlinecite{nomura2018three}, nodal lines occur off the high-symmetry $Z$ point of the Brillouin zone. While the electronic band structure displays direct band gap along the high-symmetry lines as shown in Fig.~\ref{material}(d), close inspection throughout the entire Brillouin zone reveals that the valence and conduction bands touch each other along a pair of closed nodal lines colored in red in Fig.~\ref{material}(e). Additionally, the two topmost occupied bands form another nodal line [the orange line in Fig.~\ref{material}(e)], which pierces the red nodal lines, manifesting the proposed linking structure. Moreover, it is shown that strain can induce a topological phase transition from the nodal line semimetal phase with monopole nodal lines to a 3D weak Stiefel-Whitney insulator. The pair of the monopole nodal lines appearing near the $Z$ point fuse together and pair annihilate at the $\sim$ 3 \% of tensile strain, when strain is applied along the out-of-plane ($z$) direction to 3D ABC graphdiyne with the rest of the lattice parameters fixed at the values obtained without strain. It is found that in the resulting insulator, any 2D slice in the momentum space with fixed $k_z$ has $w_2=1$, which confirms the 3D weak Stiefel-Whitney insulator phase of 3D graphdiyne under a tensile-strain.

Let us briefly mention the influence of time-reversal symmetry breaking and spin-orbit coupling. When time-reversal symmetry is broken due to effective Zeeman effect, a monopole nodal line semimetal turns into an axion insulator with quantized magnetoelectric polarizability as shown in Ref.~\onlinecite{ahn2018band}. On the other hand, when spin-orbit coupling is not negligible, a monopole nodal line semimetal becomes a higher-order topological insulator with helical hinge states~\cite{wang2018higher}. 
These examples clearly show that monopole nodal line semimetal materials can be considered as a parent state leading to various novel topological insulators under suitable conditions. 

\section{Stiefel-Whitney insulators in 2D and 3D}

Up to now, we have considered $w_2$ defined on a sphere or a torus enclosing a nodal line, which is embedded in 3D momentum space. In this section, we consider $w_2$ as a topological invariant that classifies 2D $I_{\text{ST}}$-symmetric insulators, in which $w_2$ is defined on the entire 2D Brillouin zone torus. 

\subsection{Second Stiefel-Whitney number on a torus}

2D $I_{\text{ST}}$-symmetric insulators can be characterized by the Wilson loop spectrum on the 2D Brillouin zone. The 2D Brillouin zone can be viewed as a 2D torus, parametrized by two periodic cycles $(\phi,\theta)=(k_x,k_y)$ along which occupied states may be non-orientable, contrary to the case on a sphere where occupied states are always orientable. If the Wilson loop operator is calculated along a non-orientable cycle, its spectrum cannot reveal the topological property due to the possible flat spectra existing on the $\Theta=0$ and $\Theta=\pi$ lines. For this reason, we only consider Wilson loop operators calculated along the orientable cycles below. More subtle issues related with the non-orientability is discussed in detail in Ref.~\onlinecite{ahn2018band}

\begin{figure}[t!]
\includegraphics[width=8.5cm]{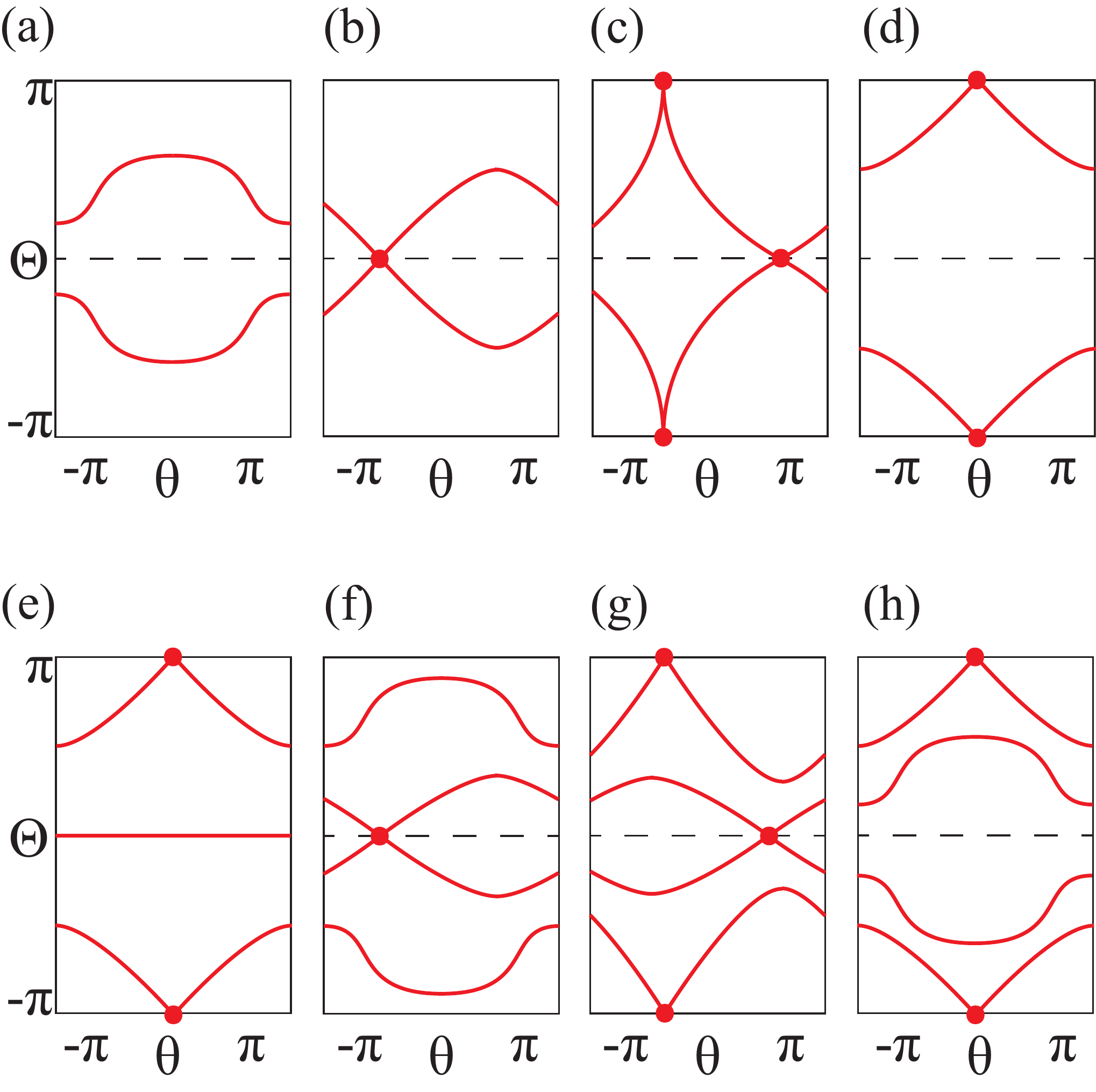}
\caption{Wilson loop spectra on a torus.
The Wilson loop operator is calculated along the orientable $\phi$ cycle at a fixed $\theta$.
(a-d) Spectrum when $N_{\text{occ}}=2$.
$(w_{1,\theta},w_2)$=
(a) $(0,0)$,
(b) $(1,0)$,
(c) $(0,1)$,
(d) $(1,1)$.
(e-h) Nontrivial spectrum when $N_{\text{occ}}=3$ and $N_{\text{occ}}=4$.
(e) When $N_{\text{occ}}=3$ with $w_2=1$. 
When the number of occupied bands is odd, $w_2$ can be determined only if we calculate the Wilson loop operator along a orientable cycle.
(f-h) When $N_{\text{occ}}=4$.
$(w_{1,\theta},w_2)$=
(f) $(1,0)$,
(g) $(0,1)$,
(h) $(1,1)$.
Here, $w_{1,\theta}$ is the first Stiefel-Whitney number computed along the $\theta$ direction.
Figures are adoped from the supplemental materials in Ref.~\onlinecite{ahn2018band}.
}
\label{Wilson_torus}
\end{figure}

Fig.~\ref{Wilson_torus} shows the Wilson loop spectra computed on a 2D torus.
As discussed before, the second Stiefel-Whitney number on a torus indicates whether the Wilson loop operator can be continuously deformed to the identity operator or not, modulo an even number of winding on non-contractible cycles.
Accordingly, the parity of the number of crossing points on $\Theta=\pi$ gives the second Stiefel-Whitney number as it does on a sphere.
For example, $w_2=0$ in Fig.~\ref{Wilson_torus}(a,b,f), and $w_2=1$ in Fig.~\ref{Wilson_torus}(c,d,e,g,h).

What makes the Wilson loop spectrum on a torus distinct from that on a sphere is the boundary condition of the Wilson loop operator. While $W=1$ at $\theta=0$ and $\pi$ on a sphere, the periodic boundary condition in Eq.~(\ref{Wilson_p.c.}) should be satisfied on a torus.
Because the boundary condition on a torus does not require that all eigenvalues are degenerate at the end-points on $\Theta=0$, an odd number of the crossing points on $\Theta=\pi$ does not necessarily mean that the eigenvalues wind as shown in Fig.~\ref{Wilson_torus}(d,e,g,h).
Moreover, when $N_{\text{occ}}$ is even, crossing points on $\Theta=0$ are protected not only locally but also globally.
As crossing points are protected on both $\Theta=0$ and $\pi$, there are three distinct topological phases characterized by an odd number of crossing points i) only on $\Theta=0$, ii) only on $\Theta=\pi$, and iii) on both $\Theta=0$ and $\Theta=\pi$.
On the other hand, when $N_{\text{occ}}$ is an odd integer, the crossing points on $\Theta=0$ are not protected due to the flat spectrum.

When $N_{\text{occ}}$ is even, there are four distinct types of Wilson loop spectra.
For instance, Fig.\ref{Wilson_torus}(a-d) correspond to the $N_{\text{occ}}=2$ case.
Because a crossing point on $\Theta=0$ is topologically stable, the spectrum in Fig.~\ref{Wilson_torus}(a) and (b) [(c) and (d)] is distinct although $w_2=0$ [$w_2=1$] in both cases. In fact, they can be distinguished by the first Stiefel-Whitney number along $\theta$ ($w_{1,\theta}$). To understand this, let us recall the periodic boundary condition for the Wilson loop operator shown in Eq.~(\ref{Wilson_p.c.}), 
\begin{align}
\label{Wilson_winding_torus}
W(2\pi)
&=M^{-1}W(0)M,
\end{align}
where $\det M=-1$ when $w_{1,\theta}=1$, in the parallel-transport gauge defined before. 
When $N_{\text{occ}}=2$ and $\det M=-1$, it becomes
\begin{align}
\exp(i\Theta(2\pi)\sigma_y)
&=M^{-1}\exp(i\Theta(0)\sigma_y)M\notag\\
&=\exp(-i\Theta(0)\sigma_y),
\end{align}
which shows that eigenvalues are interchanged as $\theta$ goes from $0$ to $2\pi$ such that an odd number of crossing points occur.
As a Wilson loop operator can be diagonalized into $2\times 2$ blocks, this applies to any case with even $N_{\text{occ}}$.
For instance, three distinct topological phases when $N_{\text{occ}}=4$ are shown in Fig.~\ref{Wilson_torus}(f,g,h) which corresponds to $(w_{1,\theta},w_2)=(1,0), (0,1)$, and $(1,1)$, respectively.


It is worth noting that $w_2$ and the corresponding Wilson loop spectrum may change depending on the unit cell choice when $w_{1,\theta}=1$.
Notice that the spectrum in Fig.~\ref{Wilson_torus}(b) and (d) differ by a constant shift  by $\pi$ while they have different second Stiefel-Whitney numbers.
The same is true for Fig.~\ref{Wilson_torus}(f) and (h).
To understand the origin of such unit cell dependence, let us use $(k_x,k_y)$ to parametrize the Brillouin zone. Since the eigenstates of the Wilson loop operator calculated along $k_x$ are Wannier states localized in the $x$-direction and the Wilson eigenvalues are Wannier centers, Fig.~\ref{Wilson_torus}(b) and (d) (also (f) and (h)) indicate that a uniform shift of the Wannier centers changes the second Stiefel-Whitney number. In other words, the second Stiefel-Whitney number can be changed if the unit cell is shifted by a half lattice constant. Therefore, $w_2$ becomes a well-defined topological invariant only when $w_1=0$. The insulator with $w_2=1$ and $w_1=0$ can be called a 2D Stiefel-Whitney insulator, which is an example of higher order topological insulators. In particular, the 2D Stiefel-Whitney insulator has fragile band topology when $N_{\text{occ}}=2$ as explained in the following section.

\subsection{Second Stiefel-Whitney number when $N_{\text{occ}}=2$: Euler class, fragile topology, and corner charges}

Although the general integral form of $w_2$ is not known, an integral form of $w_2$ can be found in some special cases. In particular, when $N_{\text{occ}}=2$ and the occupied bands are orientable, $w_2$ is identical to the parity of the Euler invariant $e_2$. The Euler invariant $e_2$ is an integer topological invariant for two real bands which can be written as a simple flux integral form~\cite{hatcher2003vector,zhao2017pt,nakahara2003geometry},
\begin{align}
\label{eq.Euler}
e_2=\frac{1}{2\pi}\oint_{BZ}d{\bf S}\cdot\tilde{\bf F}_{12},
\end{align}
where $
\tilde{\bf F}_{mn}({\bf k})
=\nabla_{\bf k}\times \tilde{\bf A}_{mn}({\bf k})
$
and
$
\tilde{\bf A}_{mn}({\bf k})
=\braket{\tilde{u}_m({\bf k})|\nabla_{\bf k}|\tilde{u}_n({\bf k})}
$
($m,n=1,2$) are $2\times 2$ antisymmetric real Berry curvature and connection defined by real states $\ket{\tilde{u}_n({\bf k})}$.
It is invariant under any $SO(2)$ gauge transformation, which has the form $O({\bf k})=\exp [-i\sigma_y\phi({\bf k})]$ and satisfies $\det[O({\bf k})]=1$.
On the other hand, under an orientation-reversing transformation with $\det[O({\bf k})]=-1$, which has the form $O({\bf k})=\sigma_z\exp [-i\sigma_y\phi({\bf k})]$,
$e_{2}$ changes its sign.
Therefore, the Euler class is well-defined only for orientable real states, that is, the states associated only with $O({\bf k})$ with a unit determinant.

The flux integral form of $e_2$ can be connected to transition functions in the following way.
To show this relation, let us notice that the 2D Brillouin zone can be deformed to a sphere when the real states are orientable along any non-contractible one-dimensional cycles as far as the topology of the real states is concerned. Then the sphere can be divided into two hemispheres, the northern ($N$) and southern ($S$) hemispheres, which overlap along the equator. Along the equator, the real smooth wave functions 
$|\tilde{u}^{N}\rangle$ and $|\tilde{u}^{S}\rangle$ defined on the northern and southern hemispheres, respectively can be connected 
by a transition function $t^{NS}=\braket{\tilde{u}^N|\tilde{u}^S}=\exp [-i\sigma_y\phi_{NS}]\in SO(2)$.
It is straightforward to show that
\begin{align}\label{eqn:e2_transition}
e_2
&=\frac{1}{2\pi}\oint_{S^2} d{\bf S}\cdot \tilde{\bf F}_{12}\notag\\
&=\frac{1}{2\pi}\int_{N} d{\bf S}\cdot \tilde{\bf F}_{12}+\frac{1}{2\pi}\int_{S} d{\bf S}\cdot \tilde{\bf F}_{12}\notag\\
&=\frac{1}{2\pi}\oint_{S^1}d{\bf k}\cdot \left(\tilde{\bf A}_{N,12}- \tilde{\bf A}_{S,12}\right)\notag\\
&=\frac{1}{2\pi}\oint_{S^1}d{\bf k}\cdot \nabla_{\bf k}\phi_{NS},
\end{align}
where $S^1$ indicates the circle along the equator.
Therefore the Euler class $e_{2}$ is identical to the winding number of the transition function $t^{NS}$. Let us note that Eq.~(\ref{eqn:e2_transition}) is also equivalent to the definition of the monopole charge module two, and thus its parity is equivalent to $w_2$ according to the discussion in Sec.IV.A. 

One physical consequence resulting from a nonzero Euler invariant $e_2$ is the existence of anomalous corner charges. The presence of corner charges can be understood in terms of the effective Hamiltonian for boundary states~\cite{wang2018higher}. Here let us briefly explain the idea. Suppose that a two-dimensional system is composed of two quantum Hall insulators with Chern numbers $c_1=1$ and $c_1=-1$, respectively, which are related to each other by $I_{\text{ST}}$. This system is an Euler insulator with $e_2=1$, which can be confirmed by the winding pattern of the Wilson loop spectrum. In this particular limit of the Euler insulator, two counter-propagating chiral edge states exist. The edge states are fully gapped after two $I_{\text{ST}}$-symmetric mass terms $m_1$ and $m_2$ are added. Each of the two mass terms has $4N_{i=1,2}+2$ zeros due to the $I_{\text{ST}}$ symmetry condition $m_{1,2}(\theta)=-m_{1,2}(-\theta)$, where $\theta$ denotes the angular coordinate of the circular boundary of a disk-shaped finite-size system, and $N_{i=1,2}$ are non-negative integers. The band gap of the edge spectrum $2m=2\sqrt{m_{1}^2+m_2^2}$ is nonzero because $m_1$ and $m_2$ do not vanish simultaneously in general. However, when there is additional chiral symmetry, only one mass term, which we take here as $m_1$, remains, so the edge band gap closes at $4N_1+2$ points. As the points of zero mass are domain kinks, charges are localized there. The corner charges are robust because they are energetically isolated from the bulk bands. Even when chiral symmetry is broken, the corner charges remain localized as long as they are in the bulk gap.

The band topology associated with the nonzero Euler class is fragile. Namely, the Wannier obstruction of an Euler insulator with $e_2\neq0$ disappears after additional trivial bands are introduced below the Fermi level~\cite{ahn2018failure}. Although the Euler class is defined only for two band systems, its parity still remains meaningful even after additional trivial bands are introduced. Namely, if the Euler class of the two-band model is even (odd), $w_{2}$ of the system should remain zero (one) after the inclusion of additional trivial bands~\cite{ahn2018band}.
Such a change of the topological indices from $Z$ to $Z_{2}$ can also be observed from the variation of the winding pattern in the Wilson loop spectrum when additional trivial bands are added~\cite{ahn2018band,bouhon2018wilson}.In fact, such fragility of the winding pattern in the Wilson loop spectrum reflects the fragility of the Wannier obstruction~\cite{cano2018topology,wang2018higher,ahn2018band,bouhon2018wilson}.
Although the nontrivial second Stiefel-Whitney number ($w_2=1$) does not induce a Wannier obstruction when the number of bands is bigger than two, anomalous corner states can still exist. Here the corner charges are induced by the configuration of the Wannier centers constrained by the non-trivial second Stiefel-Whitney number~\cite{ahn2018failure}. 



\subsection{Topological phase transition mediated by monopole nodal line, and 3D weak Stiefel-Whitney insulator}

Since the second Stiefel-Whitney number determines not only the $Z_2$ monopole charge of a nodal line on its wrapping sphere but also the $Z_2$ topological invariant of a 2D Stiefel-Whitney insulator, an intriguing topological phase transition mediated by monopole nodal lines can occur in 3D $PT$-symmetric systems. To describe this, let us start with a sphere wrapping a monopole nodal line in momentum space, and deform it into two parallel 2D planes with fixed $k_z$. Then each 2D plane can be considered as a 2D subsystem with $PT$ symmetry. Since the monopole charge of the nodal line is identical to the difference of $w_2$ of these two planes, if $w_2=0$ in one plane, $w_2=1$ in the other plane. Armed with this information, let us start from a 3D normal insulator and assume that a pair of monopole nodal line is created at the $\Gamma$ point by tuning a parameter $M$. Then every 2D subspace with fixed $k_z$ has $w_2=1$ when its $k_z$ lies between the monopole nodal line pair whereas the other 2D subspaces with $k_z$ on the other side of the Brillouin zone should have $w_2=0$. After the two monopole nodal lines are pair annihilated at the Brillouin zone boundary, one can expect that every 2D slice of the Brillouin zone with fixed $k_z$ has $w_2=1$, which is the definition of a 3D weak Stiefel-Whitney insulator.
This phase can be considered as a vertical stack of weakly interacting 2D Stiefel-Whitney insulators.
In general, a 3D weak Stiefel-Whitney insulator is characterized by three Stiefel-Whitney numbers defined on $k_x=\pi$, $k_y=\pi$, and $k_z=\pi$ planes, respectively. 
The three invariants encode the three stacking directions of 2D subsystems.
The invariants can be changed only when monopole nodal line pairs are created and then annihilated at the Brillouin zone boundary after traversing the full Brillouin zone, which is analogous to the topological phase transition between a 3D Chern insulator and a normal insulator mediated by Weyl points~\cite{ramamurthy2015patterns}.

\begin{figure}[t!]
\includegraphics[width=8.5cm]{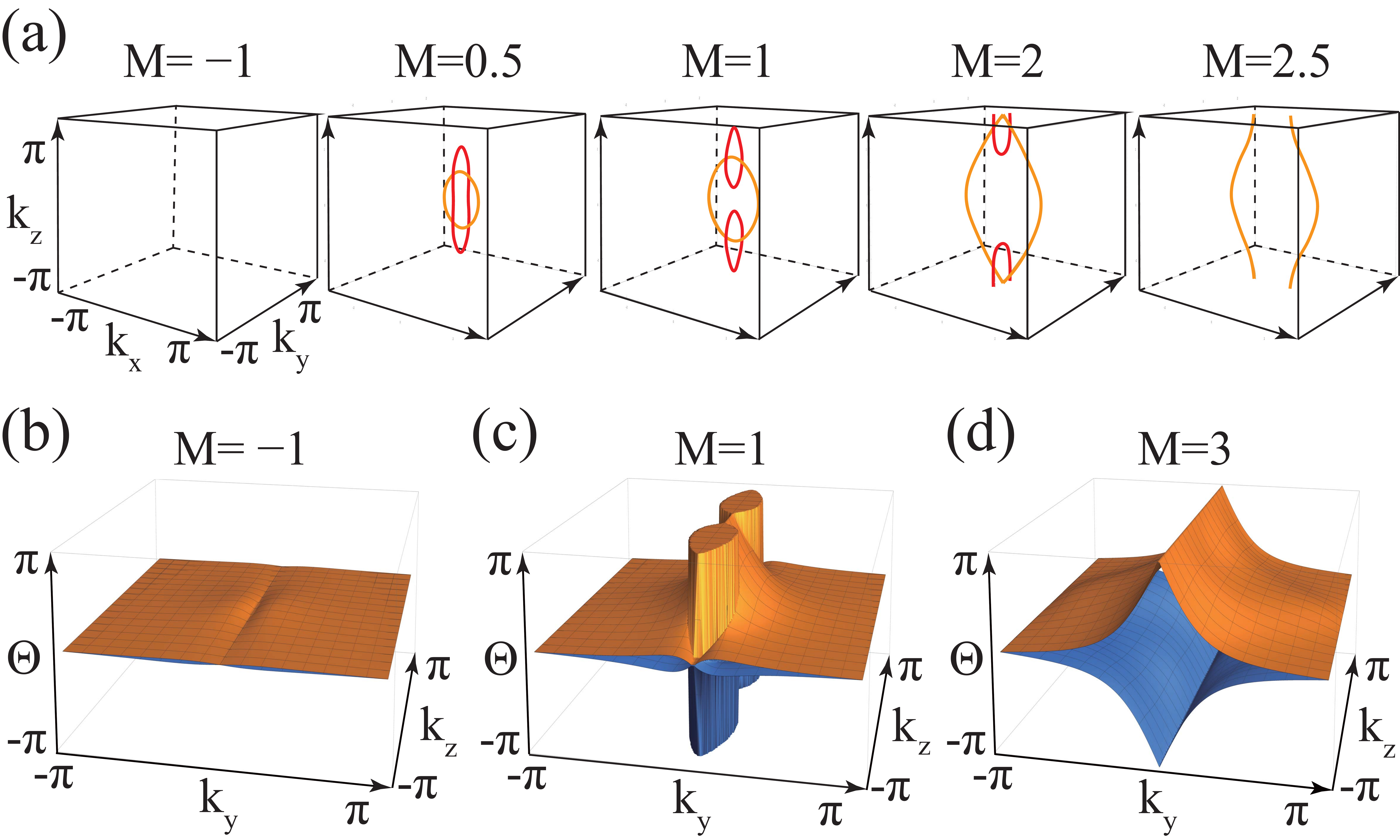}
\caption{
Topological phase transition from a normal insulator to a 3D weak Stiefel-Whitney insulator.
The Hamiltonian in Eq.~(\ref{model}) is used with $r=0.5$ and $m=0.9$.
(a) Shape of the nodal lines formed by touching between a conduction and a valence band (red) and between valence bands (orange).
As $M$ increases from $-1$ to $2.5$, a pair of monopole nodal lines are created near $(0,0,0)$, and then they are annihilated near $(0,0,\pi)$.
(b)$\sim$(d) Wilson loop operators are calculated along the $k_x$ direction at each value of $k_y$ and $k_z$.
(b) Normal insulator at $M=-1$. 
(c) Nodal line semimetal with monopole nodal lines at $M=1$. 
(d) 3D weak Stiefel-Whitney insulator at $M=3$.
}
\label{fig:TPT}
\end{figure}

To demonstrate the topological phase transition, let us consider a lattice regularization of Eq. (\ref{nlsm}),
\begin{align}
\label{model}
H({\bf k})
&=\sum_{i=1}^3 f_i({\bf k})\Gamma_i+f_{15}({\bf k})\Gamma_{15},
\end{align}
where $f_1=2\sin k_x$, $f_2=2\sin k_y$, $f_3=M+2(\cos k_x-1)+2(\cos k_y-1)+2r(\cos k_z-1)$, $f_{15}=m$, and  $r$ and $m$ are positive constants.
The energy spectrum has a simple analytic form $E=\pm\sqrt{f_1^2+\left(f_\rho\pm |m|\right)^2}$, where $f_{\rho}=\sqrt{f_2^2+f_3^2}$.
When $r<1+m/2$, an insulator-semimetal-insulator transition occurs as $M$ is varied.
If we focus on $r<1+m/2$ and $M<4-m$, the system is a normal insulator when $M<-m$, a 3D weak Stiefel-Whitney insulator when $4r+m<M<4-m$, and it is a semimetal having two monopole nodal lines when $-m<M<4r+m$.
One can clearly see that two monopole nodal lines are linked by the line of valence band degeneracy [See Fig.~\ref{fig:TPT}(a)].
The monopole nodal lines are pair-created at $(k_x,k_y,k_z)=(0,0,0)$ and pair-annihilated at $(k_x,k_y,k_z)=(0,0,\pi)$ via a double band inversion as $M$ increases from $-m$ to $4r+m$.
To verify the change of topological properties, the corresponding Wilson loop spectra are shown in Fig.~\ref{fig:TPT}(b-d).

\subsection{3D strong Stiefel-Whitney insulator and quantized magnetoelectric response}
\begin{figure}[b!]
\includegraphics[width=8.5cm]{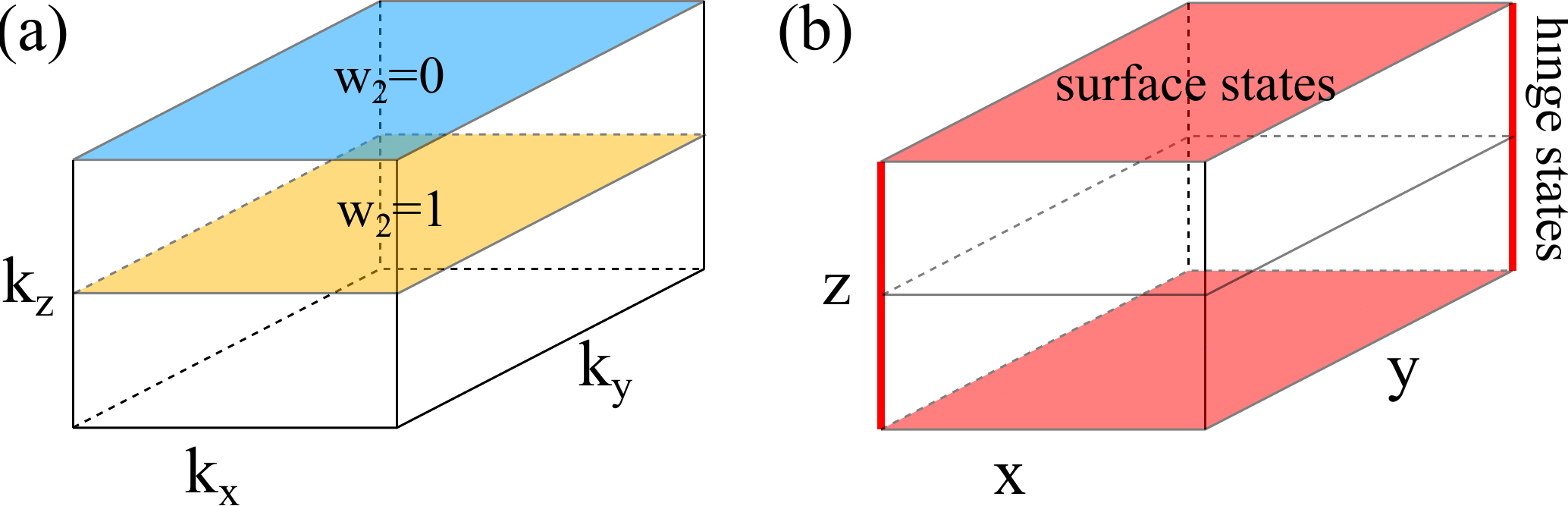}
\caption{3D strong Stiefel-Whitney insulator protected by $C_{2z}T$ symmetry.
(a) Schematic figure describing the second Stiefel-Whitney number on the $C_{2z}T$-invariant planes in momentum space.
In a 3D strong Stiefel-Whitney insulator, $w_2(k_z=\pi)-w(k_z=0)=1$ modulo two.
(b) Schematic figure describing the gapless states on the surface and hinges in real space.
An odd number of 2D Dirac fermions appear on each of the top and bottom surfaces.
1D chiral fermions appear on the side hinges.
Figures are adoped from Ref.~\onlinecite{ahn2018higher}.
}
\label{fig:3DstrongStiefel-Whitney insulator}
\end{figure}
Since we have a new 2D $Z_2$ invariant $w_2$ and the associated 2D $Z_2$ topological insulator (2D Stiefel-Whitney insulator), it is natural to ask whether one can find a 3D topological insulator associated with the second Stiefel Whitney class. In spinless fermionic systems with $I_{\text{ST}}=PT$, real wave functions can be defined over the full 3D Brillouin zone. However, unfortunately, there is no corresponding 3D topological invariant~\cite{zhao2016unified,bzdusek2017robust}. So we do not expect a 3D topological insulator associated with the Stiefel-Whitney number other than the 3D weak Stiefel-Whitney insulator discussed above. Instead, we focus on the 3D systems with $I_{\text{ST}}=C_{2z}T$ where the $z$-axis is chosen as the axis for $C_2$ rotation. In $C_{2z}T$-symmetric 3D systems, only the wave functions on the $k_z=0$ and $k_{z}=\pi$ planes can be real with the corresponding second Stiefel-Whitney numbers $w_{2}(0)$ and $w_{2}(\pi)$, respectively. Thus, a 3D strong $Z_{2}$ topological invariant $\Delta_{3D}$ may be defined as $\Delta_{3D}\equiv w_{2}(\pi)-w_{2}(0)$ in a way similar to how the 3D Fu-Kane-Mele invariant is constructed. Since $\Delta_{3D}$ originates from $w_{2}$ in $I_{\text{ST}}$-invariant planes, the 3D topological insulator with $\Delta_{3D}=1$ can be called {\it a 3D strong Stiefel-Whitney insulator}. Let us note that the idea of $C_{2z}T$-protected $Z_2$ topological insulator was already proposed in Ref.~\onlinecite{fang2015new}. However, its bulk electromagnetic response and the related hinge excitations are studied recently in Ref.~\onlinecite{ahn2018higher,wieder2018pump}. In fact, a strong 3D Stiefel-Whitney insulator is an example of higher-order topological insulators whose bulk magnetoelectric response is described by the axion term with the quantized magnetoelectric polarizability $P_3=1/2$. As a result of the bulk boundary correspondence associated wth the quantized $P_3$, we show that a 3D strong Stiefel-Whitney insulator has chiral hinge states~\cite{zhang2013surface,fang2017rotation,khalaf2018higher,
kooi2018inversion,varnava2018surfaces,vanmiert2018higher,
yue2018symmetry,schindler2018higher,
ezawa2018strong,ezawa2018magnetic} along the edges parallel to the rotation axis and 2D massless Dirac fermions on the surfaces normal to the rotation axis as shown in Fig.~\ref{fig:3DstrongStiefel-Whitney insulator}.

The equivalence between $\Delta_{3D}$ and the quantized magnetoelectric polarizability can be shown by analyzing the homotopy group of the sewing matrix $G$ for $I_{\text{ST}}$ symmetry defined as
\begin{align}
\label{eq.sewing}
G_{mn}({\bf k})
&=\braket{u_{m(-C_{2z}{\bf k})}|C_{2z}T|u_{n\bf k}},
\end{align}
which satisfies
\begin{align}
\label{eq.sewing_constraint}
G_{mn}({\bf k})
&=G_{nm}(-C_{2z}{\bf k}),
\end{align}
where $-C_{2z}{\bf k}=(k_x,k_y,-k_z)$ and $\ket{u_{n\bf k}}$ is the cell-periodic part of a Bloch state. If we choose smooth occupied states, the corresponding sewing matrix also becomes smooth. Then the nontrivial homotopy class of $G$ characterizes the obstruction to taking a uniform representation $G({\bf k})=G_0$ independent of ${\bf k}$.
At a generic momentum, $G({\bf k})\in U(N)$. On the other hand, on a $C_{2z}T$-invariant plane, either the $k_z=0$ or $k_z=\pi$ plane, $G^T({\bf k})=G({\bf k})$ according to Eq.~\eqref{eq.sewing_constraint}, from which we obtain 
\begin{align}
G({\bf k})\in U(N)/SO(N),
\end{align}
where $N$ denotes the number of occupied bands.

In a smooth complex gauge,
the magnetoelectric polarizability $P_3$ takes the form of the 3D Chern-Simons invariant~\cite{qi2008topological,wang2010equivalent}
\begin{align}
P_3
&=\frac{1}{8\pi^2}\int_{\rm BZ}d^3k\epsilon^{ijk}{\rm Tr}\left[A_i\d_jA_k-\frac{2i}{3}A_iA_jA_k\right],
\end{align}
where $A_{mn}({\bf k})=\braket{u_{m\bf k}|i\nabla_{\bf k}|u_{n\bf k}}$ is the Berry connection, and ${\rm BZ}$ denotes the Brillouin zone. In terms of the sewing matrix $G({\bf k})$, one can show that
\begin{align}
2P_3=\frac{1}{24\pi^2}\int_{\rm BZ}d^3k\epsilon^{ijk}{\rm Tr}\left[(G^{-1}\d_iG)(G^{-1}\d_jG)(G^{-1}\d_kG)\right],
\end{align}
which is nothing but the 3D winding number of the sewing matrix $G$.
Since the 3D winding number is determined by the 2D winding numbers in invariant planes, we eventually find
\begin{align}
\label{strong=P3}
2P_3=w_2(\pi)+w_2(0)=\Delta_{\rm 3D} (\text{ mod 2}),
\end{align}
which is proved more explicitly in Ref.~\onlinecite{ahn2018higher}.

The relation between the bulk topological invariant of the axion insulator and that of the 2D Stiefel-Whitney insulator implies a similar relation between their anomalous boundary states. In fact, the chiral hinge states in an axion insulator can be considered to result from the pumping of charges at the corners of the Stiefel-Whitney insulator when $k_z$ is regarded as a parameter for the pumping process~\cite{wieder2018pump}. This charge pumping picture can be extended further to the cases with strong spin-orbit coupling. Here one can consider a helical charge pumping, where the corner charges with different spins are pumped to the opposite directions, which leads to a construction of a higher-order topological insulator with helical hinge states~\cite{fang2017rotation,song2018diagnosis,wang2018higher}.

\section{Discussion}
The Stiefel-Whitney classes are examples of characteristic classes, which are the cohomology classes associated to vector bundles, describing how the corresponding vector bundle is twisted~\cite{hatcher2003vector,hatcher2002algebraic}. There are mainly four different types of characteristic classes known up to now: Chern classes, Stiefel Whitney classes, Euler Classes, and Pontryagin classes~\cite{hatcher2003vector,hatcher2002algebraic}. While the idea of Chern classes and associated topological invariants, such as Chern numbers, mirror or spin Chern numbers, Fu-Kane invariants, have been widely applied to condensed matter physics, the implication of the other characteristic classes in the context of condensed matter physics is not well established yet. In this paper, we have reviewed the recent progress in the study of topological physics associated with Stiefel-Whitney numbers. More explicitly, we showed that the first Stiefel-Whitney number is equivalent to the quantized Berry phase so that the nontrivial first Stiefel-Whitney number $w_1$ indicates either a 1D insulator with quantized charge polarization or a stable Dirac point (nodal line) in 2D Dirac semimetals (3D nodal line semimetals) in systems with space-time inversion symmetry $I_{\text{ST}}$. Moreover, we proved that the second Stiefel-Whitney number not only characterizes the monopole charge of nodal lines in $I_{\text{ST}}$-symmetric systems but also serves as a well-defined 2D topological invariant characterizing a 2D Stiefel-Whitney insulator. This idea is further extended to 3D systems with $I_{\text{ST}}=PT$ and $I_{\text{ST}}=C_{2z}T$ leading to the 3D weak and strong Stiefel-Whitney insulators, respectively. However, materials that realize 2D Stiefel-Whitney insulator and 3D strong Stiefel-Whitney insulator composed of spinless fermions are still lacking, which provides new research opportunities to find novel topological materials and phenomena. 

Also, as briefly explained, $I_{\text{ST}}$-symmetric two-band systems in 2D can be characterized by another characteristic class, so-called the Euler class. The Euler class is an integer topological invariant classifying real orientable two-band systems. As discussed before, a two-band system with a nonzero Euler invariant $e_2$ has fragile band topology and supports corner charges. In fact, an Euler insulator with $e_2\neq0$ has various intriguing topological properties which are not discussed in this review article. For instance, a two-band system with the Euler invariant $e_{2}$ always possesses band crossing points whose total winding number is equal to $2e_2$~\cite{ahn2018failure}. Thus the conventional Nielsen-Ninomiya theorem fails in systems with a nonzero Euler invariant. Moreover, if an additional trivial band is coupled to the original two-band system and band crossing happens between them, the newly generated Dirac points play the role of the source of $\pi$ Berry phase with a Dirac string in between, which strongly affects the braiding properties of the original Dirac points. Such a nontrivial braiding is a manifestation of the non-abelian topological charge of real wave functions, which is discussed thoroughly in Ref.~\onlinecite{wu2018beyond,ahn2018failure}.  To unveil novel topological physics associated with other characteristic classes is definitely one important issue for future research. 

\begin{acknowledgments}
{\it Acknowledgment.|}
J.A. and S.P. were supported by IBS-R009-D1.
B.-J.Y. was supported by the Institute for Basic Science in Korea (Grant No. IBS-R009-D1) and Basic Science Research Program through the National Research Foundation of Korea (NRF) (Grant No.0426-20180011), and  the POSCO Science Fellowship of POSCO TJ Park Foundation (No.0426-20180002).
This work was supported in part by the U.S. Army Research Office under Grant Number W911NF-18-1-0137.
Y.K. was supported by Institute for Basic Science (IBS-R011-D1) and NRF grant funded by the Korea government (MSIP) (NRF-2017R1G1B5018169). 
D. K. was supported by Samsung Science and Technology Foundation under Project Number SSTF-BA1701-07 and Basic Science Research Program through NRF funded by the Ministry of Education (NRF-2018R1A6A3A11044335). The computational calculations were performed using the resource of Korea institute of Science and technology information (KISTI).
\end{acknowledgments}


\providecommand{\newblock}{}

\end{document}